\documentclass[runningheads]{svmult}

\usepackage{makeidx}   % allows index generation
\usepackage{graphicx}  % standard LaTeX graphics tool
\usepackage{subeqnar}  % subnumbers individual equations
\usepackage{multicol}  % used for the two-column index
\usepackage{physprbb}  % modified textarea for proceedings,
\usepackage{latexsym}  % add math symbols
\makeindex             % used for the subject index
\newcommand{\greeksym}[1]{{\usefont{U}{psy}{m}{n}#1}}
\newcommand{\upr}[1]{\mbox{\greeksym{#1}}}

\newcommand{\bfit}[1]{\textit{\textbf{#1}\/}}

\newcommand{\be}{\begin{equation}}
\newcommand{\ee}{\end{equation}}
\newcommand{\bi}{\begin{itemize}}
\newcommand{\ei}{\end{itemize}}
\newcommand{\lf}[2]{\mbox{\large $\frac{#1}{#2}$}}

\newcommand{\bOx}{\raisebox{-1.2 pt}{\Large $\Box$}}

\begin{document}

\title*{The detection of Gravitational Waves}

\toctitle{The detection of Gravitational Waves}
% allows explicit linebreak for the table of content
%

\titlerunning{The detection of Gravitational Waves}
% allows abbreviation of title, if the full title is too long
% to fit in the running head

\author{J.\ Alberto Lobo \inst{1}}
%\and Elsa Bertino\inst{2}}
%
\authorrunning{J.\ Alberto Lobo}
% if there are more than two authors,
% please abbreviate author list for running head
%
\institute{Departament de F\'\i sica Fonamental, Diagonal 647,
	   E-08028 Barcelona, Spain}

% \and Universit\'{e} de Paris-Sud,
%     Laboratoire d'Analyse Num\'{e}rique,
%     B\^{a}timent 425,\\
%     F-91405 Orsay Cedex, France}

\maketitle

\begin{abstract}
This chapter is concerned with the question: how do gravitational
waves (GWs) interact with their detectors? It is intended to be a
\emph{theoretical review} of the fundamental concepts involved in
interferometric and acoustic (Weber bar) GW antennas. In particular,
the type of signal the GW deposits in the detector in each case
will be assessed, as well as its intensity and deconvolution.
Brief reference will also be made to detector sensitivity
characterisation, including very summary data on current
state of the art GW detectors.
\end{abstract}

\section{Introduction}
\label{sec.1}

Gravitational waves (GW), on very general grounds, appear to be a
largely unavoidable consequence of a well established fact that no
known interaction propagates instantly from source to observer:
gravitation would be the first exception to this rule, should it
be described by Newton's theory. Indeed, the Newtonian gravitational
potential $\phi(\vec{x},t)$ satisfies Poisson's equation

\begin{equation}
 \nabla^2\phi(\vec{x},t) = -4\pi G\,\varrho(\vec{x},t)\;,
 \label{eq1.1}
\end{equation}
where $\varrho(\vec{x},t)$ is the density of matter in the sources
of gravitational fields, and $G\/$ is Newton's constant. But,
since~(\ref{eq1.1}) contains no time derivatives, the time dependence
of $\phi(\vec{x},t)$ is purely \emph{parametric}, i.e., time variations
in $\varrho(\vec{x},t)$ \emph{instantly} carry over to $\phi(\vec{x},t)$,
irrespective of the value of $\vec{x}$. So, for example, non-spherically
symmetric fluctuations in the mass distribution of the Sun (such as e.g.
those caused by solar storms) would instantly \emph{and} simultaneously
be felt both in the nearby Mercury and in the remote Pluto\ldots

Quite independently of the \emph{quantitative} relevance of such
instant propagation effect in this particular example -- which
is none in practice --, its very existence is \emph{conceptually}
distressing. In addition, the asymmetry between the space and
time variables in (\ref{eq1.1}) does not even comply with the
basic requirements of Special Relativity.

Einstein's solution to the problem of gravity, General Relativity
(GR), does indeed \emph{predict} the existence of radiation of
gravitational waves. As early as 1918, Einstein himself provided a
full description of the polarisation and propagation properties of
weak GWs~\cite{ein18}. According to GR, GWs travel across otherwise
flat empty space at the speed of light, and have \emph{two} independent
and \emph{transverse} polarisation amplitudes, often denoted
$h_+(\vec{x},t)$ and $h_\times(\vec{x},t)$, respectively~\cite{mtw}.
In a more general framework of so called \emph{metric theories} of
gravity, GWs are allowed to have up to a maximum 6 amplitudes, some
of them transverse and some longitudinal \cite{will}.

The theoretically predicted existence of GWs poses of course the
experimental challenge to \emph{measure} them. Historically, it took
a long while even to attempt the construction of a \emph{gravitational
telescope}: it was not until the decade of the 1960's that J.\ Weber
first took up the initiative, and developed the first gravitational
antennas. These were elastic cylinders of aluminum, most sensitive
to \emph{short bursts} (a few milliseconds) of GWs. After analysing
the data generated by \emph{two} independent instruments, and looking
for events in coincidence in both, he reported evidence that a
considerably large number of GW flares had been sighted~\cite{weber}.

Even though Weber never gave up his claims of \emph{real} GW
detection~\cite{jweb2}, his contentions eventually proved untenable.
For example, the rate and intensity of the reported events would
imply the happening of several supernova explosions per week in
our galaxy~\cite{gh71}, which is astrophysically very unlikely.

It became clear that more sensitive detectors were necessary, whose
design and development began shortly afterwards. In the mid 1970's and
early eighties, the new concept \emph{interferometric} GW detector
started to develop~\cite{weiss,ron}, which would later lead to the larger
\emph{LIGO} and \emph{VIRGO} projects, as well as others of more reduced
dimensions (\emph{GEO-600} and \emph{TAMA}), and to the future \emph{space}
antenna \emph{LISA}. In parallel, \emph{cryogenic} resonant detectors were
designed and constructed in several laboratories, and towards mid 1990's
the next generation of \emph{ultracryogenic} antennas, \emph{NAUTILUS}
(Rome), \emph{AURIGA} (Padua), \emph{ALLEGRO} (Baton Rouge, Louisiana)
and \emph{NIOBE} (Australia), began \emph{taking data}. More recently,
data exchange protocols have been signed up for multiple detector
coincidence analysis~\cite{igec}. Based on analogous physical principles,
new generation \emph{spherical} GW detectors are being programmed in
Brazil, Holland and \mbox{Italy~\cite{msch,mgrail,sfera}.}

In spite of many years of endeavours and hard work, GWs have proved
elusive to all dedicated detectors constructed so far. However, the
discovery of the binary pulsar {\sl PSR 1913+16\/} by R.\ Hulse and
J.\ Taylor in 1974~\cite{ht75}, and the subsequent long term detailed
monitoring of its orbital motion, brought a breeze of fresh air into
GW science: the measured decay of the orbital period of the binary
system due to \emph{gravitational bremsstrahlung} accurately conforms
to the predictions of General Relativity. Hulse and Taylor were
awarded the 1993 Nobel Prize in Physics for their remarkable work.
As of 1994~\cite{tay94}, the accumulated binary pulsar data confirm
GR to a high precision of a tenth of a percent\footnote{
It appears that priorities in pulsar observations have since shifted
to other topics of astrophysical interest, so it is difficult to find
more recent information on {\sl PSR~1913+16}.}.

The binary pulsar certainly provides the most compelling evidence to date
of the GW phenomenon as such, yet it does so thanks to the observation of
a \emph{back action} effect on the source. Even though I do not consider
accurate the statement, at times made by various people, that this is
only \emph{indirect} evidence of GWs, it is definitely a matter of fact
that there is more to GWs than revealed by the binary pulsar\ldots\ For
example, amplitude, phase and polarisation parameters of a GW can only
be measured, according to current lore, with dedicated GW antennas.

%it also constitutes a beautiful example of
%a \emph{clean} source of gravitational radiation. By this it is usually
%meant a system whose dynamics can be theoretically established, at least
%in principle, to arbitrary precision -- whence it can be understood in full
%physical detail. Other astrophysical objects, such as e.g.\ supernovae, are
%much more difficult to understand, because certain essential ingredients
%are largely unknown, for instance the equations of state, etc. -- they are
%not so clean. The GW signal from a binary system can be reliably modeled
%as a function of only a few parameters~\cite{f3m}, and it actually
%constitutes one of the soundest known candidates for detection by future
%antennas~\cite{cofo,thorne}.

But how do GW telescopes interact with the radiation they are supposed
to detect? This is of course a fundamental question, and is also the
subject of the present contribution, where I intend to review the
\emph{theoretical foundations} of this problem, and its solutions as
presently understood. In section~\ref{sec.2}, I summarise the essential
properties of GWs within a rather large class of possible theories of
the gravitational interaction; section~\ref{sec.3} briefly bridges
the way to sections~\ref{sec.4} and~\ref{sec.5} where interferometric
and acoustic detector concepts are respectively analysed in some detail.
Section~\ref{sec.5} also includes aboundant reference to new generation
spherical detectors in its various variants (solid, hollow, dual). For
the sake of completeness, I have added a section (section~\ref{sec.6})
with a very short summary of detector characterisation concepts, so that
the interested reader gets a flavour of how sensitivities are defined,
what do they express and how do GW signals compare with local noise
disturbances in currently conceived detectors. Section~\ref{sec.7}
closes the paper with a few general remarks.

\section{The nature of gravitational waves}
\label{sec.2}

Quite generally, a time varying mass-energy distribution creates in
its surroundings a time varying gravitational field (\emph{curvature}).
As already stressed in section~\ref{sec.1}, we do not expect such time
variations to travel instantly to distant places, but rather that they
travel as \emph{gravity waves} across the intervening space.

Now, how are these waves ``seen''? A \emph{single} observer may of course
not feel any variations in the gravitational field where he/she is immersed,
if he/she is in \emph{free fall} in that field -- this is a consequence of
the \emph{Equivalence Principle}~\cite{wein}. \emph{Two} nearby observers
have instead the capacity to do so: for, both being in free fall, they can
take each other as a reference to measure any relative \emph{accelerations}
caused by a non-uniform gravitational field, in particular those caused
by a \emph{gravitational wave field}. We can rephrase this argument
saying that gravitational waves show up as \emph{local tides}, or
gradients of the local gravitational field at the observatory.

In the language of Differential Geometry, tides are identified as
\emph{geodesic deviations}, i.e., variations in the four-vector
connecting nearby geodesic lines. It is shown in textbooks,
e.g.~\cite{mtw}, that the geodesic deviation equation is

\begin{equation}
 \frac{D^2\xi_\mu}{ds^2} = R_{\mu\nu\rho\sigma}\,\dot x^\nu\dot x^\sigma
 \xi^\rho\; ,
 \label{eq2.1}
\end{equation}
where ``$D\/$'' means \emph{covariant derivative}, $s\/$ is proper
time for \emph{either} geodesic, $R_{\mu\nu\rho\sigma}$ is the Riemann
tensor, $\dot x^\nu$ is a unit tangent vector (again to either geodesic),
and~$\xi^\mu\/$ is the vector connecting corresponding points of the
two geodesic lines.

The GW fields considered in this paper will be restricted to a class of
\emph{perturbations} of the geometry of otherwise flat space-time, with
the additional assumptions that they be

\begin{center}
\begin{tabular}{l}
-- small \\
-- time-dependent \\
-- vacuum perturbations
\end{tabular}
\end{center}

This is certainly not the most general definition of a GW yet it will
suffice to our purposes here: any GWs generated in astrophysical
sources and reaching a man-made detector definitely satisfy the
above requirements. The interested reader is referred to~\cite{grif}
for a thorough treatment of more general GWs.

Following the above assumptions, a GW can be described by perturbations
$h_{\mu\nu}(\vec{x},t)$ of a \emph{flat} Lorentzian metric
$\eta_{\mu\nu}\/$, i.e., there exist \emph{quasi-Lorentzian}
coordinates $(\vec{x},t)$ in which the space-time metric
$g_{\mu\nu}(\vec{x},t)$ can be written

\begin{equation}
 g_{\mu\nu}(\vec{x},t) = \eta_{\mu\nu} + h_{\mu\nu}(\vec{x},t)
 \label{eq2.2}
\end{equation}
with

\begin{equation}
 \eta_{\mu\nu} = {\rm diag}\;(-1,1,1,1)\ ,\qquad
 \left|h_{\mu\nu}(\vec{x},t)\right|\ll 1\;.
 \label{eq2.2.5}
\end{equation}

The actual effect of a GW on a pair of test particles is, according
to~(\ref{eq2.1}), determined by the Riemann tensor $R_{\mu\nu\rho\sigma}$,
and this in turn is determined by the functions $h_{\mu\nu}\/$. I now
review briefly the different possibilities in terms of which is the
theory underlying the physics of gravity waves, i.e., which are the
field equations which the $h_{\mu\nu}\/$ satisfy.

\subsection{Plane GWs according to General Relativity}
\label{sub2.1}

The vacuum field equations of General Relativity are, as is well
known~\cite{mtw},

\begin{equation}
 R_{\mu\nu} = 0\;,
 \label{eq2.3}
\end{equation}
where $R_{\mu\nu}\/$ is the Ricci tensor of the metric $g_{\mu\nu}$.
If quadratic and successively higher order terms in the perturbations
$h_{\mu\nu}\/$ are neglected then this tensor can be seen to be given by

\begin{equation}
 R_{\mu\nu} = \bOx\bar h_{\mu\nu} - \partial_\rho\partial_\nu\bar h_\mu^\nu
  - \partial_\mu\partial_\nu\bar h_\rho^\nu\;,
 \label{eq2.4}
\end{equation}
with

\begin{equation}
 \bOx\equiv\eta^{\mu\nu}\,\partial_\mu\partial_\nu\ ,\qquad
 \bar h_{\mu\nu}\equiv h_{\mu\nu} - \lf{1}{2}\,h\,\eta_{\mu\nu}\ ,\qquad
 h\equiv\eta^{\mu\nu}\,h_{\mu\nu}\;.
 \label{eq2.5}
\end{equation}

New coordinates $(\vec{x'},t')$ can be defined by means of transformation
equations

\begin{equation}
 x'^\mu = x^\mu + \varepsilon^\mu(\vec{x},t)\;,
 \label{eq2.6}
\end{equation}
and these will still be quasi-Lorentzian if the functions
$\varepsilon^\mu(\vec{x},t)$ are \emph{sufficiently small}.
More precisely, the GW components are, in the new coordinates,

\begin{equation}
 h'_{\mu\nu} = h_{\mu\nu} - \partial_\mu\varepsilon_\nu
 - \partial_\nu\varepsilon_\mu
 \label{eq2.7}
\end{equation}
provided higher order terms in $\varepsilon^\mu$ are neglected. Thus
``sufficiently small'' means that the derivatives of $\varepsilon_\mu$
be of the order of magnitude of the metric perturbations $h_{\mu\nu}$
-- so that in the new coordinates $x'^\mu\/$ the metric tensor
\emph{also} splits up as $g'_{\mu\nu}=\eta_{\mu\nu}+h'_{\mu\nu}$.
It is now possible, see~\cite{mtw}, to choose new coordinates in
such a way that the \emph{gauge conditions}

\begin{equation}
 \partial_\nu\bar h_\rho^\nu = 0
 \label{eq2.8}
\end{equation}
hold. This being the case, equations~(\ref{eq2.3}) read

\begin{equation}
 \bOx\bar h_{\mu\nu} = 0\;,
 \label{eq2.9}
\end{equation}
which are vacuum \emph{wave equations}. Therefore GWs travel across
empty space at the speed of light, according to GR theory. \emph{Plane
wave solutions} to (\ref{eq2.9}) satisfying (\ref{eq2.8}) can now
be constructed~\cite{mtw} which take the form

\begin{equation}
 h_{\mu\nu}(\vec{x},t) = h_{\mu\nu}^{TT}(\vec{x},t) = \left(
 \begin{array}{cccc}
  0 & 0 & 0 & 0 \\
  0 &   h_+(t-z)        &  h_\times(t-z) & 0 \\
  0 & \ h_\times(t-z)\  & \ -h_+(t-z)\   & 0 \\
  0 & 0 & 0 & 0
 \end{array}\right)
 \label{eq2.10}
\end{equation}
for waves travelling down the $z\/$-axis. The label $TT\/$ is an
acronym for \emph{transverse-traceless}, the usual denomination
for this particular gauge.

The physical meaning of the polarisation amplitudes in (\ref{eq2.10})
is clarified by looking at the effect of an incoming wave on test
particles. Consider e.g.\ two equal test masses whose center of mass
is at the origin of $TT\/$ coordinates; let $\ell_0$ be their distance
in the absence of GWs, and $(\theta,\varphi)$ the orientation (relative
to the $TT\/$ axes) of the vector joining both masses. Making use of
the geodesic deviation equation~(\ref{eq2.1}), with the Riemann tensor
associated to~(\ref{eq2.10}), it can be seen that the GW only affects
the \emph{transverse} projection of the distance relative to the wave
propagation direction (the $z\/$-axis); in fact, if
$\ell(t)\equiv(\xi^\mu\,\xi_\mu)^{1/2}$ is such distance then some simple
algebra leads to the result\footnote{
Note that the Riemann tensor is calculated at the center of mass of the
test particles, therefore at $\vec{x}=0$. But it can also be calculated
at the position of \emph{either} one of them~--~this would only make up
for a negligible \emph{second order} difference.}

\begin{equation}
 \ell(t) = \ell_0\,\left[1 + \frac{1}{2}\,\left\{
 h_+(t)\,\cos 2\varphi + h_\times(t)\,\sin 2\varphi\right\}\,
 \sin^2\theta\right]\;.
 \label{eq2.12}
\end{equation}

\begin{figure}[t]
\centering
\includegraphics[width=11.9cm]{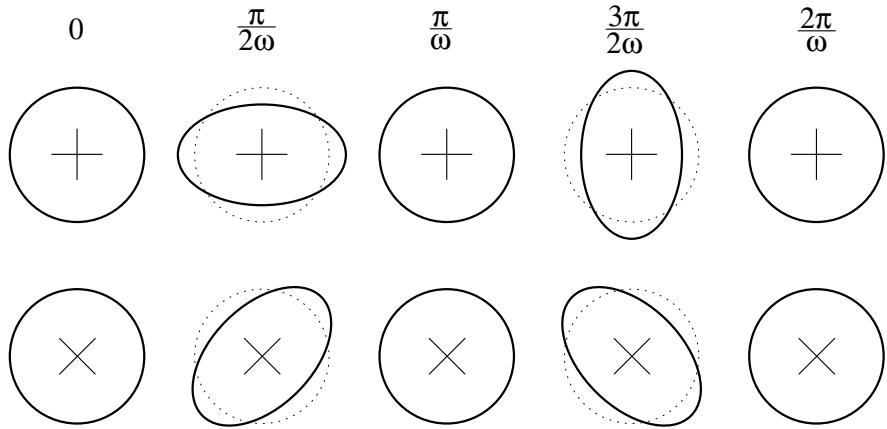}
\caption{The ``$+$'' and ``$\times$'' polarisation modes of a GW,
according to GR theory. The dotted lines (circles) indicate the test
particles' position in the absence of GW signal. Each step in the graph
corresponds to a quarter of the period of the driving GW, as labeled atop.
\label{fig1}}
\end{figure}

It is very important at this point to stress that the \emph{wavelength}
$\lambda$ of the incoming GW must be much larger than the distance
$\ell(t)$ between the particles for~(\ref{eq2.12}) to hold, i.e.,

\begin{equation}
 \ell(t)\ll\lambda\;,
 \label{eq2.13}
\end{equation}
and this is a condition which must be \emph{added} to the already made
assumption that $|h_{\mu\nu}|\ll 1$.

A graphical representation of the result (\ref{eq2.12}) is displayed in
figure~\ref{fig1}: a number of test particles are evenly distributed
around a circle perpendicular to the incoming GW, i.e., in the $xy\/$
plane. When a periodic signal comes in, the distances between those
particles change following~(\ref{eq2.12}); note that the changes are
modulated by the angular factors, i.e., according to the particles'
positions on the circle. The ``$+$'' mode is characterised by a
vanishing wave amplitude $h_\times$, while in the ``$\times$'' mode
$h_+$ vanishes.

\subsection{Plane GWs according to \bfit{metric theories} of gravity}
\label{sub2.2}

Although General Relativity has never been questioned so far by
experiment, there are in fact alternative theories, e.g.\ Brans-Dicke
theory~\cite{bd61}, which are interesting for a number of reasons,
for example \emph{cosmological} reasons~\cite{gazta}. Generally,
these theories make their own specific predictions about GWs, and
they partly differ from those of GR -- just discussed. The term
\emph{metric theory} indicates that the gravitational interaction
affects the \emph{geometry} of space-time, i.e., the metric tensor
$g_{\mu\nu}$ is a fundamental ingredient -- though other fields may
also be necessary to complete the theoretical scheme. Obviously,
General Relativity falls within this class of theories.

The appropriate scheme to assess the physics of such more general class
of GWs was provided long ago~\cite{el73}. The idea is to consider only
\emph{plane gravitational waves}, which should be an extremely good
approximation for astrophysics, given our great distance even to the
nearest conceivable GW source, and to characterise them by their
\emph{Newman-Penrose scalars}~\cite{chan}.

It appears that only \emph{six} components of the Riemann tensor out
of the usual~20 are independent in a \emph{plane} GW; these are given
by the four Newman-Penrose scalars

\begin{equation}
 \upr{Y}_2(v), \upr{Y}_3(v), \upr{Y}_4(v)\ \; {\rm y}\ \;\upr{F}_{22}(v)\;,
 \label{eq2.14}
\end{equation}
of which $\upr{Y}_2$ and $\upr{F}_{22}$ are real, while $\upr{Y}_3$
and $\upr{Y}_4$ are \emph{complex} functions of the null variable $v\/$
-- see~\cite{chan} for all notation details. If a quasi-Lorentzian
coordinate system is chosen such that GWs travel along the $z\/$-axis
then $v=t-z$, and one can calculate the scalars~(\ref{eq2.14}) to obtain

\begin{subeqnarray}
 \upr{Y}_2(t-z) & = & -\lf{1}{6}\,R_{tztz}\;, \label{eq2.15a} \\
 \upr{Y}_3(t-z) & = & -\lf{1}{2}\,(-R_{txtz}+\I R_{tytz})\;, \label{eq2.15b} \\
 \upr{Y}_4(t-z) & = & -R_{txtx}+R_{tyty}+2\I R_{txty}\;, \label{eq2.15c} \\
 \upr{F}_{22}(t-z) & = & -R_{txtx}-R_{tyty}\;. \label{eq2.15d}
\end{subeqnarray}

General Relativity is characterised by $\upr{Y}_4(t-z)$ being the only
non-vanishing scalar, while in Brans-Dicke theory $\upr{F}_{22}(t-z)$
also is different from zero -- see~\cite{el73} for full details.

It is relevant to remark at this stage that the only non-trivial
components of the Riemann tensor of a plane GW are the so called
``\emph{electric}'' components, $R_{titj}$, as we see in
equations~(\ref{eq2.15a}--d) above. These are, incidentally,
\emph{also} the only ones which appear in the geodesic deviation
equation~(\ref{eq2.1}), since one may naturally choose
$\dot x^\mu=(1;0,0,0)$ \footnote{
Note that, with this choice, $R_{\mu\rho\nu\sigma}\,\dot x^\mu\dot x^\nu
=R_{t\rho t\sigma}$ but, because of the symmetries of the Riemann tensor,
only values of $\rho$ and $\sigma$ different from zero, i.e.,
$\left\{\rho\sigma\right\} = \left\{ij\right\}$, give non-zero
contributions.}.

\begin{figure}[t]
\centering
\includegraphics[width=11cm]{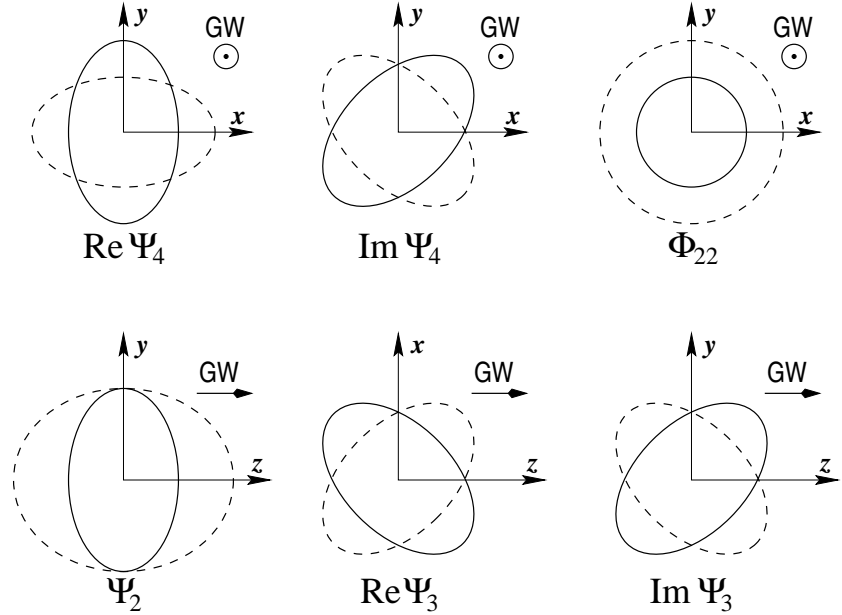}
\caption{The six polarisation modes of a plane, metric GW. The dotted
and solid lines correspond to distributions separated by a half period
of the incoming wave, and it is assumed that the particles lie on a
circle when no GW is acting. Note the indicators that the modes in the
upper row are excited by a wave which arrives \emph{perpendicular} to
the particles' plane, while those in the lower row correspond to GWs
which are in the \emph{same} plane as the particles. It must however
be clarified that the mode $\upr{F}_{22}$ has spherical symmetry, so it
includes a combination of longitudinal and transverse excitations in
like proportions.
\label{fig2}}
\end{figure}

This fact helps us make a graphical representation of all \emph{six}
possible polarisation states of a general metric GW in the same manner
as in figure~\ref{fig1}. The result is represented in figure~\ref{fig2}
-- whose source is reference~\cite{will}. The idea is to take a ring
of test particles, let a GW pass by, and analyse the results of the
displacements it causes in the distributions of those particles, just
as done in section~\ref{sub2.2}. Note that the first three modes are
\emph{transverse}, while the other three are longitudinal -- see the
caption to the figure. As already stressed, GR only gives rise to the
two $\upr{Y}_4$ modes.

\section{Gravitational wave detection concepts}
\label{sec.3}

We are now ready to discuss the objectives and procedures to detect
GWs: knowing their physical structure, one can design systems whose
interaction with the GWs be sufficiently well understood and under
control; suitable monitoring of the dynamics of such systems will
be the source of information on whatever GW parameters may show
up in a given observation experiment.

There are \emph{two} major detection concepts: \emph{interferometric}
and \emph{acoustic} detection. Historically, the latter came first
through the pioneering work of J.\ Weber, but interferometric GW
antennas are at present attracting the larger stake of the investment
in this research field, both in hardware and in human commitment. This
is because much hope has been deposited in their capabilities to reach
sufficient sensitivity to \emph{measure} GWs for the first time.

Interferometric detectors aim to measure phase shifts between light rays
shone along two different (straight) lines, whose ends are defined by
\emph{freely suspended} test masses. This is done in a Michelson layout,
using mirrors, beam splitters and photodiodes. Acoustic detectors are
instead based on elastically linked test masses -- rather than freely
suspended -- which \emph{resonantly} respond to GW excitations.

These \emph{qualitative} ideas can be made quantitatively precise, but
the process is a non-trivial one and has important subtleties which must
be properly understood for a thorough assessment of the detector workings
and readout. The next sections are devoted to explain with some detail
which are the \emph{theoretical principles} governing the behaviour and
response of both kinds of GW antennas.

\section{\bfit{Interferometric} GW detectors}
\label{sec.4}

\begin{figure}[t]
\centering
\includegraphics[width=11cm]{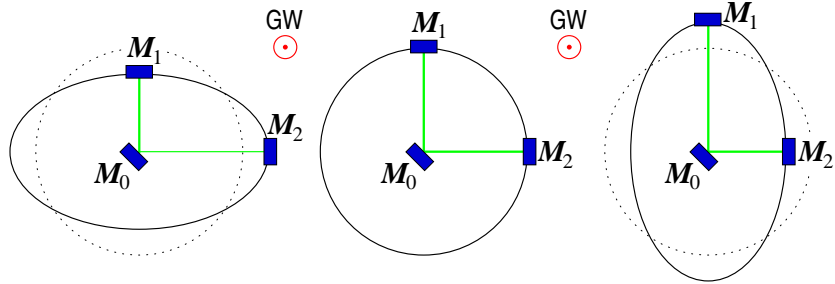}
\caption{The ``na\"\i ve'' interferometric detector concept: a GW coming
perpendicular to the sheet's plane, ``$+$'' polarised relative to the
$x\/$ and $y\/$ axes, causes the end masses $M_1$ and $M_2$ to oscillate
in phase opposition relative to the central mass $M_0$. Light is shone
into the system, and suitable beam splitters and mirrors are attached
to the masses; length changes are then measured interferometrically,
which directly lead to determine the GW amplitude.
\label{fig3}}
\end{figure}

A rather \emph{na\"\i ve} idea to measure the effect of an incoming GW
is provided by the following argument -- see also figure~\ref{fig3} for
reference: let a GW having a ``$+$'' polarisation (assume GR for simplicity
at this stage) come in perpendicular to the local horizontal at a given
observatory; if three masses are laid down on the vertices of an ideally
oriented isosceles right triangle then, as we saw (figure~\ref{fig1}),
the catets shrink and stretch with opposite phases. If a beam of laser
light is now shone into the system, and a beam-splitter attached to
mass $M_0$ and mirrors attached to $M_1$ and $M_2$, then one can
think of measuring the distance changes between the masses by simple
interferometry.

This may look like a very reasonable proposal for a detector yet the
following criticism readily suggests itself: gravitation is concerned
with \emph{geometry}, i.e., gravitational fields alter lengths and
angles; therefore GWs will affect \emph{identically} both distances
between the masses of figure~\ref{fig3} \emph{and} the wavelength of
the light travelling between them -- thus leading to a \emph{cancellation}
of the conjectured interferometric effect\ldots

While the criticism is certainly correct, the conclusion is \emph{not}.
The reason is that it overlooks the fact that gravitation is concerned
with the geometry of \emph{space-time} -- not just space. In the case
of the above GW it so happens that, in $TT\/$ coordinates, the time
dimension of space-time is not \emph{warped} in the GW geometry
-- see the form of the metric tensor in equation~(\ref{eq2.10}) --
while the transverse space dimensions \emph{are}. Consequently an
electromagnetic wave travelling in the $xy\/$ plane experiences
\emph{wavelength} changes depending on the propagation direction,
but it does \emph{not} experience \emph{frequency} changes. The
net result of this is that the \emph{phase} of the electromagnetic
wave differs from direction to direction of the $xy\/$ plane, and
\emph{this} makes a GW amenable to detection by interferometric
principles.

Looked at in this way, the masses represented in figure~\ref{fig3}
only play a \emph{passive} role in the detector, in the sense that
they simply make the interference between the two light beams possible
by providing physical support for the mirrors and beam splitters. In
other words, the physical principles underlying the working of an
interferometric GW detector must have to do with the \emph{interaction
between GWs and electromagnetic waves} rather than with geodesic
deviations of the \emph{masses}.

Admittedly, this is not the most common point of view~\cite{brian}. It
can however be made precise by the following considerations, which are
studied in depth in references~\cite{cqg} and~\cite{fara}.

\subsection{Test light beams in a GW-warped space-time}
\label{sub3.2}

According to the above, it appears that we must address the question of
how GW-induced fluctuations in the geometry of a background space-time
affect the properties (amplitude and phase) of a plane electromagnetic
wave -- a \emph{light beam}~-- which travels through such space-time.
It will be sufficient to consider that the light beam is a \emph{test}
beam, i.e., its back action on the surrounding geometry is negligible.

Let then $A_\mu(\vec{x},t)$ be the vector potential which describes
an electromagnetic wave travelling in vacuum; $A_\mu\/$ thus satisfies 
Maxwell's equations

\begin{equation}
 \bOx A_\mu = 0\;,
 \label{eq3.1}
\end{equation}
where $\bOx$ stands for the generalised d'Alembert operator:

\begin{equation}
 \bOx\equiv g^{\rho\sigma}\,\nabla_\rho\nabla_\sigma\;.
 \label{eq3.2}
\end{equation}

We need only retain \emph{first order} terms in $h_{\mu\nu}$ in
the covariant derivatives, so that equation~(\ref{eq3.1}) reads

\begin{equation}
\bOx A_\mu\simeq\eta^{\rho\sigma}\,(\partial_\rho\partial_\sigma A_\mu
 - 2\Gamma_{\mu\sigma}^\nu\,\partial_\rho A_\nu
 - \Gamma_{\rho\sigma}^\nu\,\partial_\nu A_\mu
 - A_\nu\,\partial_\rho\Gamma_{\mu\sigma}^\nu) = 0\;,
 \label{eq3.3}
\end{equation}
To this equation, \emph{gauge conditions} must be added. We shall
conventionally adopt the usual Lorentz conditions,
$\nabla_{\!\!\mu} A^\mu=0$, which, to lowest order in the gravitational
perturbations, read

\begin{equation}
 \nabla_{\!\mu} A^\mu\simeq
 -\partial_t A_t + \partial_x A_x + \partial_y A_y + \partial_z A_z -
 h^{\mu\nu}\,\partial_\mu A_\nu = 0\;.
 \label{eq3.4}
\end{equation}

In addition to the weakness of the GW perturbation, it is also the case
in actual practice that:

\begin{itemize}
\item The GW typical frequencies, $\omega$, are much smaller than the
 frequency of the light, $\Omega$: $\omega\ll\Omega$.
\item The wavelength of the GW is much larger than the cross sectional
 dimensions of the light beam.
\end{itemize}

Wave front distortions and beam curvature are effects which can also
be safely neglected in first order calculations~\cite{fara}. Finally,
I shall make the simplifying assumption of perpendicular incidence,
i.e., the incoming GW propagates in a direction orthogonal to the
interferometer's arms\footnote{
This condition can be easily relaxed, but it complicates the equations
to an extent which is inconvenient for the purposes of the present
review. Details are fully given in reference~\protect\cite{cqg}.}.

We shall thus consider one of the interferometer arms aligned with
the $x\/$-direction, and the other with the $y\/$-direction, while
the incoming GW will be assumed to approach the detector down the
$z\/-axis$. The interaction GW-light beam will thus occur in the
$z=0$ plane, hence the GW perturbations can be suitably described
by a function of time alone, i.e.,

\begin{equation}
 h_{\mu\nu}(\vec{x},t)\longrightarrow h_{\mu\nu}(\omega t)\;,
 \label{eq3.5}
\end{equation}
where $\omega$ is the frequency of the GW, which we can also safely
assume to be plane-fronted, since its source will in all cases of
interest be far removed from the observatory. In addition, for a
beam running along the $x-axis$, the electromagnetic vector potential
will only depend on the space variable $x\/$ and on time,~i.e.,

\begin{equation}
 A_{\mu}(\vec{x},t)\longrightarrow A_{\mu}(x,t)\;.
 \label{eq3.6}
\end{equation}

The following \emph{ansatz} suggests itself for a solution to
equations~(\ref{eq3.3}):

\begin{subeqnarray}
 A_t\,(t,x) & = & \varepsilon_0(\omega t)\;, \label{eq3.7a} \\
 A_x\,(t,x) & = & \varepsilon_1(\omega t)\;, \label{eq3.7b} \\
 A_y\,(t,x) & = & a_2\;\exp[{\rm i}\Omega\,(t-x)]\,
		{\rm e}^{{\rm i}\,\phi_2(\omega t)}\;, \label{eq3.7c} \\
 A_z\,(t,x) & = & a_3\;\exp[{\rm i}\Omega\,(t-x)]\,
		{\rm e}^{{\rm i}\,\phi_3(\omega t)}\;, \label{eq3.7d}
\end{subeqnarray}
where $\varepsilon_0(\omega t)$, $\varepsilon_1(\omega t)$, $\phi_2(\omega t)$
and $\phi_3(\omega t)$ are small quantities of order $h\/$, $a_2$ and
$a_3$ are constants, and $\Omega$ is the frequency of the light. Clearly,
these expressions reproduce the plane wave solutions to vacuum Maxwell's
equations in the limit of flat space-time, i.e., when $h_{\mu\nu}=0$.

Let us now take an incoming GW of the form

\begin{equation}
 h_{\{+,\times\}}(t) = \tilde H_{\{+,\times\}}\,\E^{\I\omega t}\;,
 \label{eq3.8}
\end{equation}
and substitute it into (\ref{eq3.3}) and (\ref{eq3.4}), with the
\emph{ansatz}~(\ref{eq3.7a}-d), neglecting higher order terms in
the ratio $\omega/\Omega$. Then~\cite{cqg}, \emph{both} $\phi_2$
and $\phi_3$ are seen to satisfy the approximate differential equation

\begin{equation}
 \ddot\phi(t) + \frac{2\I\Omega}{\omega}\,\dot\phi(t) +
 \frac{\I\Omega^2}{\omega^2}\,h_+(t) = 0\;.
 \label{eq3.9}
\end{equation}

The solution to this equation which is \emph{independent of the initial
conditions} is, to the stated level of accuracy,

\begin{equation}
 \phi_2(t)\simeq\phi_3(t)\simeq\frac{\Omega}{2\omega}\,
 \tilde H_+\sin\omega t\;.
 \label{eq3.10}
\end{equation}

As shown in reference~\cite{cqg}, we need not worry about either
$\varepsilon_0(\omega t)$ or $\varepsilon_1(\omega t)$ at this stage
because the \emph{longitudinal} component of the electric field (i.e.,
$E_x\/$) is an order of approximation smaller than the transverse
components, which are given by

\begin{equation}
 E_y=\partial_tA_y\ ,\qquad E_z=\partial_tA_z\;,
 \label{eq3.11}
\end{equation}
hence

\begin{subeqnarray}
 E_y(x,t) & \simeq & \I\Omega\,a_2\,\exp\left[\I\Omega(t-x) +
	\I\,\frac{\Omega}{2\omega}\,\tilde H_+\sin\omega t\right]\;,
	\label{eq3.12a} \\[1 ex]
 E_z(x,t) & \simeq & \I\Omega\,a_3\,\exp\left[\I\Omega(t-x) +
	\I\,\frac{\Omega}{2\omega}\,\tilde H_+\sin\omega t\right]\;.
	\label{eq3.12b}
\end{subeqnarray}

These expressions beautifully show how the incoming GW causes
a \emph{phase shift} in an electromagnetic beam of light. Note that
this phase shift is a \emph{periodic} function of time, with the
frequency of the GW. If we consider a real interferometer, such as
very schematically shown in figure~\ref{fig3}, and call $\tau\/$ the
\emph{round trip} time for the light to go from $M_0$ to $M_2$ and
back, then the accumulated phase shift is, according to these formulas,

\begin{equation}
 \delta_x\phi = 2\times\frac{\Omega}{2\omega}\,
 \tilde H_+\sin\frac{\omega\tau}{2}\;,
 \label{eq3.13}
\end{equation}
since there is an obvious symmetry between light rays travelling to the
right and to the left for a GW arriving perpendicularly to them\footnote{
The reader is warned that this symmetry does \emph{not} happen if the GW
and the light beam are not perpendicular, see~\protect\cite{fara}.}.

The arguments leading to equations (\ref{eq3.12a}-b) can be very easily
reproduced, \emph{mutatis mutandi}, to obtain the phase shift experienced
by a light ray travelling in the $y$, rather than the $x\/$ direction
-- everything in fact amounts to a simple interchange
$\{x\longleftrightarrow y\}$ in the equations, which includes
$\{h_+\longleftrightarrow -h_+\}$ as this is equivalent to
$\{h_{xx}\longleftrightarrow h_{yy}\}$, see~(\ref{eq2.10}). The result is

\begin{equation}
 \delta_y\phi = -2\times\frac{\Omega}{2\omega}\,
 \tilde H_+\sin\frac{\omega\tau}{2}\;.
 \label{eq3.14}
\end{equation}

In the actual interferometer, provided it has equal arm lengths, the two
laser rays recombine in the beam splitter with a net phase difference

\begin{equation}
 \delta\phi = \delta_x\phi-\delta_y\phi = \frac{2\Omega}{\omega}\,
 \tilde H_+\sin\frac{\omega\tau}{2}\;,
 \label{eq3.15}
\end{equation}
and this produces an \emph{interference signal}, which is in principle
measurable -- if the instrumentation is sufficiently sensitive.

The reader may wonder how is it that the detector signal only depends
on one of the GW amplitudes, $h_+$, but not on the other, $h_\times$.
The reason is that we have made a very special assumption regarding
the orientation of the GW's polarisation axes relative to the light
beam propagation directions. In a realistic case, even if perpendicular
GW incidence happens, the detector's arms will not be aligned with the
GW's natural axes, let alone the most likely case of \emph{oblique}
incidence. An important conclusion one should draw from this section
is a \emph{conceptual} one, that interferometric detectors are able
to measure GW amplitudes and polarisations as a result of the
interaction between the electromagnetic field of light rays
and the background space-time geometry they travel across.

Beyond this, though, equation~(\ref{eq3.15}) has very relevant
\emph{quantitative} consequences, too. For example, as stressed in
reference~\cite{cqg}, its range of validity is \emph{not} limited to
interferometer arms short compared to the GW wavelength. Therefore,
according to the formula, a \emph{null} effect (signal cancellation)
happens if the round trip time $\tau$ equals the period of the GW,
$2\pi/\omega$. Likewise, equation~(\ref{eq3.15}) also tells us that
\emph{maximum} detector signal occurs when $\tau=\pi/\omega$. All this
happen to be true for arbitrary incidence and polarisation of the
incoming GWs as well. For GW frequencies in the 1\,kHz range, the best
detector should thus have arm lengths in the range of 150~kilometers
--~and even longer for lower GW frequencies. No ground based GW
antenna has ever been conceived of such dimensions yet there are
intelligent ways to \emph{store} the light in shorter arms for
suitably tuned GW periods. I shall not go into details of these
technical matters, see~\cite{ron} and~\cite{brian} for thorough
information.

\begin{figure}[b]
\centering
\includegraphics[width=7cm]{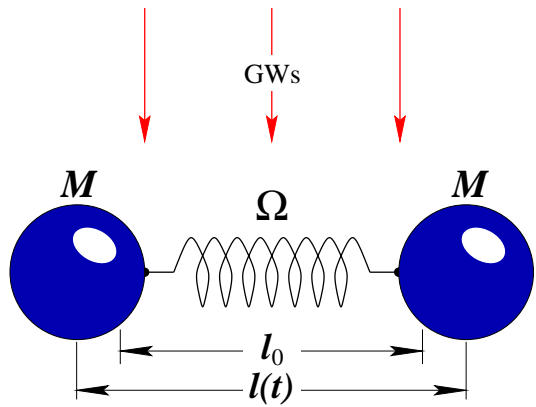}
\caption{The acoustic detector concept: a GW coming perpendicular to the
spring drives the two masses $M\/$ at its ends to oscillate at the GW's
frequency. Resonant amplification is obtained when the latter equals the
spring's characteristic frequency, $\Omega$.
\label{fig4}}
\end{figure}

\section{\bfit{Acoustic} GW detectors}
\label{sec.5}

Acoustic GW detectors work based on a completely different concept
-- see figure~\ref{fig4}: the idea is to set up test masses $M\/$
linked together by a spring of relaxed length $\ell_0$, so that GW
\emph{tides} drive their oscillations around the equilibrium position,
with significant mechanical amplification at the spring's characteristic
frequency $\Omega$. The spring deformation

\begin{equation}
 q(t)\equiv\ell(t) - \ell_0
 \label{eq5.1}
\end{equation}
thus obeys the following equation of motion\footnote{
I shall not include any \emph{dissipative} terms at this stage, for
they do not influence the key points of our present discussion.}:

\begin{equation}
 \ddot q(t) + \Omega^2 q(t) = \frac{1}{2}\,\ell_0\ddot h(t)\;,
 \label{eq5.2}
\end{equation}
where

\begin{equation}
 h(t) = \left[
 h_+(t)\,\cos 2\varphi + h_\times(t)\,\sin 2\varphi\right]\,\sin^2\theta\;,
 \label{eq5.3}
\end{equation}
as follows from (\ref{eq2.12}) -- see also reference~\cite{pizz}
for a complete discussion of this case.

This is the main idea, but in practice acoustic GW detectors are
\emph{elastic solids} rather than a single spring, i.e., they do
not have a single characteristic frequency but a whole spectrum.
The response of an elastic solid to a GW excitation is assessed
by means of the classical theory of Elasticity, as described for
example in~\cite{ll70}. In such theory the deformations of the
solid are given by the values of a vector field of displacements,
$\vec{u}(\vec{x},t)$, which satisfies the evolution equations

\begin{equation}
 \varrho\frac{\partial^2 \vec{u}}{\partial t^2} - \mu\nabla^2 \vec{u} -
 (\lambda+\mu)\,\nabla(\nabla{\bf\cdot}\vec{u}) = \vec{f}(\vec{x},t)\;,
 \label{eq5.4}
\end{equation}
where $\varrho$ is the (undeformed) solid's density, and $\lambda$ and
$\mu\/$ are its Lam\'e coefficients, related to the Poisson ratio and
Young modulus of the material the solid is made of~\cite{ll70}. The
function in the rhs of the equation is the density of \emph{external
forces} driving the motion of the system; in the present case, these
are the \emph{tides} generated by the sweeping GW, i.e.,

\begin{equation}
 f_i(\vec{x},t) = \varrho\,R_{titj}(t)\,x_j\;,
 \label{eq5.5}
\end{equation}
where $R_{titj}(t)$ are components of the Riemann tensor evaluated
at a fixed point of the solid, most expediently chosen at its center
of mass. As already discussed in section~\ref{sub2.2}, plane GWs
have at most six degrees of freedom, adequately associated with the
six \emph{electric components} of the Riemann tensor, $R_{titj}(t)$.
The six components are \emph{one monopole} amplitude and \emph{five
quadrupole} amplitudes, and this important structure is made clear
by the following expression of the density of GW tidal forces:

\begin{equation}
 \vec{f}(\vec{x},t) = \vec{f}^{(0,0)}(\vec{x})\,g^{(0,0)}(t)\ +\ 
 \sum_{m=-2}^2\,\vec{f}^{(2,m)}(\vec{x})\,g^{(2,m)}(t)\;,
 \label{eq5.6}
\end{equation}
which is entirely equivalent to (\ref{eq5.5}) -- see
reference~\cite{lobo} --, and uses the common $(l,m)$ notation
convention to indicate multipole terms. It is very important to
stress at this stage that $\vec{f}^{(l,m)}(\vec{x})$ are pure
\emph{form factors}, simply depending on the fact that tides are
monopole-quadrupole quantities, while all the relevant \emph{dynamic}
information carried by the GW is encoded in the time dependent
coefficients $g^{(l,m)}(t)$. According to these considerations,
we see that the ultimate objective of an acoustic GW antenna is
to produce values of $g^{(l,m)}(t)$ --~or indeed to extract from
the device's readout as much information as possible about those
quantities.

Somewhat lengthy algebra permits to write down a \emph{formal solution}
to equations~(\ref{eq5.5}) and~(\ref{eq5.6}) in terms of an orthogonal
series expansion~\cite{lobo}:

\begin{equation}
 \vec{u}(\vec{x},t) = \sum_N\,\omega_N^{-1}\,\vec{u}_N(\vec{x})\,
%   \sum_{\mbox{\scriptsize $\begin{array}{c}
%   l=0\ \mbox{and}\ 2 \\ m=-l,...,l \end{array}$}}
 \left[\ \sum_{\stackrel{\scriptstyle l=0\ {\rm and}\ 2}{m=-l,...,l}}\;
 f_N^{(l,m)}\,g_N^{(l,m)}(t)\,\right]\;,
 \label{eq5.7}
\end{equation}
where

\begin{subeqnarray}
 f_N^{(l,m)} & \equiv & \frac{1}{M}\,\int_{\rm Solid}\vec{u}_{N}^*(\vec{x})
 \cdot\vec{f}^{(l,m)}(\vec{x})\,\D^3x\;,	\label{eq5.8a} \\[1 ex]
 g_N^{(l,m)}(t) & \equiv & \int_0^tg^{(l,m)}(t')\,\sin\omega_N(t-t')\,\D t'\;,
 \label{eq5.8b}
\end{subeqnarray}
with $M\/$ the whole solid's mass; $omega_N\/$ is the (possibly
\emph{degenerate}) characteristic frequency of the elastic body,
and $\vec{u}_{N}(\vec{x})$ the corresponding \emph{wavefunction},
both determined by the solution to the \emph{eigenvalue} problem

\begin{equation}
 \mu\nabla^2\vec{u}_N + (\lambda+\mu)\,\nabla(\nabla\vec{\cdot}\vec{u}_N)
 = - \omega_N^2\,\varrho\,\vec{u}_N\;,
 \label{eq5.9}
\end{equation}
with the boundary conditions that the surface of the solid be free of
any tensions and/or tractions --~see~\cite{lobo} for full details.

Historically, the first GW antennas were Weber's elastic \emph{cylinders}
\cite{weber}, but more recently, \emph{spherical} detectors have been
seriously considered for the next generation of GW antennas, as they
show a number of important advantages over cylinders. I shall devote
the next sections to a discussion of both types of systems, though
clear priority will be given to spheres, due to their much richer
capabilities and theoretical interest.

\subsection{Cylinders}
\label{sub5.1}

First thing to study the response of an elastic solid to an incoming GW
is, as we have just seen, to determine its characteristic oscillation
\emph{modes}, i.e., its frequency spectrum $\omega_N$ and associated
wavefunctions, $\vec{u}_N(\vec{x})$. In the case of a cylinder this is
a formidable task; although its formal solution is known~\cite{ric,Ras},
cylindrical antennas happen in practice tobe narrow and
long~\cite{expl,naut}, and so approximate solutions can be used instead
which are much simpler to handle, and sufficiently good -- see
also~\cite{wein}.

\begin{figure}[t]
\centering
\includegraphics[width=8cm]{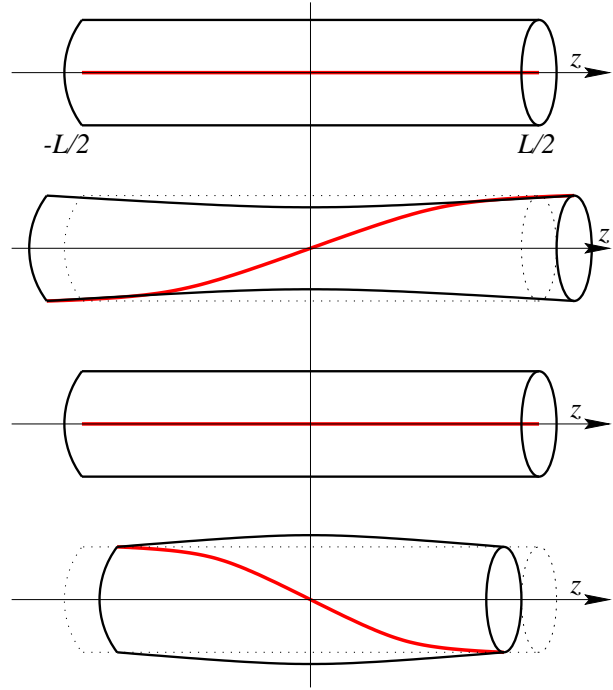}
\caption{The first longitudinal oscillation mode of a long cylinder. Note
that it has a node at the center and maxima at the end faces. A whole
period is represented, and transverse deformations are also shown.
\label{fig5}}
\end{figure}

It appears that, in the \emph{long rod} approximation, the most
efficiently coupled modes are the \emph{longitudinal} ones, and
these have typical sinusoidal profiles, of the type

\begin{equation}
 \delta z(z,t)\propto\sin\left(\frac{n\pi z}{L}\right)\,
 \sin\left(\frac{n\pi v_st}{L}\right)\,\ ,\qquad
  n=1,2,3,\ldots
 \label{eq5.10}
\end{equation}
for a rod of length $L\/$ whose end faces are at $z=\pm L/2$, and
in which the speed of sound is $v_s$. Figure~\ref{fig5} graphically
shows the longitudinal deformations of the cylinder which correspond
to~(\ref{eq5.10}), including \emph{transverse} distortions which,
though not reflected in the simplified equation~(\ref{eq5.10}), do
happen in practice as a result of the Poisson ratio being different
from zero~\cite{ll70}. An important detail to keep in mind is that
\emph{odd n} modes have maximum displacements at the end faces,
while \emph{even n} modes have nodes there. In fact, the latter do
not couple to GWs~\cite{pizz}. It is also interesting to stress that
the center of the cylinder is always a node --~this is relevant e.g.\ 
to suspension design issues~\cite{naut}.

A very useful concept to characterise the sensitivity of an acoustic
antenna is its \emph{cross section} for the absorption of GW energy.
If an incoming GW flux density of $\Phi(\omega)$ watts per square metre
and Hertz sweeps the cylinder and sets it to oscillate with energy
$E(\omega)$ joules then the cross section is \emph{defined} by the~ratio

\begin{equation}
 \sigma_{\rm abs}(\omega) = \frac{E(\omega)}{\Phi(\omega)}\;,
 \label{eq5.11}
\end{equation}
which is thus measured in m$^2$\,Hz. Simple calculations show that, for
optimal antenna orientation (perpendicular to the GW incidence direction)
this quantity is given by~\cite{pizz}

\begin{equation}
 \sigma_{\rm abs}(\omega_n) = \frac{8}{\pi n^2}\;\frac{GMv_s^2}{c^3}
 \ ,\qquad n=1,3,5,\ldots\;,
 \label{eq5.12}
\end{equation}
where $M\/$ is the cylinder's total mass.

It is interesting to get a flavour of the order of magnitude this
quantity has: consider a cylinder of Al5056 (an aluminum alloy, for
which $v_s\simeq 5400$\,m/s), 3~metres long and 60~centimetres across,
which has a mass of 2.3~tons\footnote{
These figures correspond to a real antenna, see~\protect\cite{expl}.};
the above formula tells us that, for the first mode ($n=1$),

\begin{equation}
 \sigma_{\rm abs}(\omega_1) = 4.3\times 10^{-21}\ {\rm cm}^2\,{\rm Hz}\;.
 \label{eq5.13}
\end{equation}

This is a \emph{very small} number indeed, and gives an idea of how
weak the coupling between GWs and matter is.

The weakness of the coupling gives an indication of how difficult it
is to detect GWs. By the same token, though, GWs are very weakly
damped as they travel through matter, which means they can produce
information about otherwise invisible regions, such as the interior
of a supernova, or even the big bang.

Equation~(\ref{eq5.12}) is only valid for \emph{perpendicular} GW
incidence. If incidence is instead oblique then a significant damping
factor of $\sin^4\theta\/$ comes in, where $\theta\/$ is the angle between
the GW direction and the cylinder's axis~\cite{pizz}\footnote{
Such factor can incidentally be inferred easily from~(\protect\ref{eq2.12}),
if one notices that the energy of oscillation appearing in the numerator
of~(\protect\ref{eq5.12}) is proportional to $\dot\ell^2$.}.
This is a severe penalty, and it also happens in interferometric
detectors not optimally oriented~\cite{cqg,st}.

\subsection{Solid spheres}
\label{sub5.2}

The first initiatives to construct and operate GW detectors are due
to J.\ Weber, who decided to use elastic cylinders. This philosophy
and practice has survived him\footnote{
Professor Joseph Weber died on 30th September 2000 at the age of 81.},
and still today (February 2002) all GW detectors in continuous data
taking regimes are actually Weber bars, though with significant
sensitivity improvements~\cite{bars} derived, amongst other, from
ultracryogenic and {\sl SQUID\/} techniques.

About ten years after Weber began his research, R.\ Forward published
an article~\cite{fo71} where he pondered in a semi-quantitative way
the potential virtues of a \emph{spherical}, rather than cylindrical
GW detector. Ashby and Dreitlein~\cite{ad75} estimated how the whole
Earth, as an \emph{auto-gravitating} system, responds to GWs bathing
it, and later Wagoner~\cite{wp77} developed a theoretical model to
study the response of an elastic sphere to GW excitations.

% I personally find particularly remarkable that so much effort has
% been devoted to enhance a detector, the \emph{cylinder}, some of
% whose intrinsic limitations could have been overcome with a new
% design \emph{spherical} system. It appears that the need to use
% a \emph{multiple channel} readout system --~I will explain this
% below in detail~-- was considered to pose unsurmountable difficulties.

Interest in this new theoretical concept then waned to eventually
re-emerge in the 1990s. The {\sl ALLEGRO\/} detector group at Louisiana
constructed a room temperature prototype antenna~\cite{jm93,phd}, which
produced sound experimental evidence that it is actually possible to have
a working system capable of making \emph{multimode} measurements --~I'll
come to this in detail shortly --, thus proving that a full-fledged
spherical GW detector is within reach of current technological state
of the art, as developed for Weber bars. It was apparently the fears
to find unsurmountable difficulties in this problem which deterred
further research on spherical GW antennas for years~\cite{eug}.

In this section I will give the main principles and results of the
\emph{theory} of the spherical GW detector, based on a formalism which
has already been partly used in section~\ref{sec.5}, and for whose complete
detail the reader is referred to~\cite{lobo}.

As we have seen, first thing we need is the \emph{eigenmodes} and
\emph{frequency spectrum} of the spherical solid. This is a classical
problem, long known in the literature, the solution to which I will
briefly review here, with some added emphasis on the issues of our
present concern.

The oscillation eigenmodes of a solid elastic body fall into two
families: \emph{spheroidal} and \emph{torsional} modes~\cite{lobo}.
Of these, only the former couple to GWs, while torsional modes do
not couple at all~\cite{bian}. Spheroidal wavefunctions have the
analytic form

\begin{equation}
 \vec{u}_{nlm}(\vec{x}) = A_{nl}(r)\,Y_{lm}(\vec{n})\,\vec{n}
  - B_{nl}(r)\,\I\vec{n}\!\times\!\vec{L}Y_{lm}(\vec{n})\;,
 \label{eq5.14}
\end{equation}
where $Y_{lm}$ are spherical harmonics~\cite{ed60}, $\vec{n}=\vec{x}/R\/$
is the outward pointing normal, $\vec{L}$ is the `angular momentum
operator', $\vec{L}\equiv -\I\vec{x}\times\nabla$, $A_{nl}(r)$ and
$B_{nl}(r)$ are somewhat complicated combinations of spherical Bessel
functions~\cite{lobo}, and $\{nlm\}$ are `quantum numbers' which label
the modes. The frequency spectrum appears to be composed of ascending
series of \emph{multipole harmonics}, $\omega_{nl}$, i.e., for
\emph{each} multipole value~$l\/$ there are an infinite number of
frequency harmonics, ordered by increasing values of $n$. For example,
there are \emph{monopole} frequency harmonics $\omega_{10}$, $\omega_{20}$,
$\omega_{30}$, etc.; then \emph{dipole} frequencies $\omega_{11}$,
$\omega_{21}$,\ldots, then \emph{quadrupole} harmonics $\omega_{12}$,
$\omega_{22}$, and so on. Each of these frequencies is $(2l+1)$-fold
degenerate, and this is a fundamental fact which makes of the sphere
a theoretically ideal GW detector, as we shall shortly see.

If the above expressions are substituted into (\ref{eq5.8a}-b),
then into~(\ref{eq5.7}), one easily obtains the sphere's response
function as

\begin{eqnarray}
 \vec{u}(\vec{x},t) & = & \sum_{n=1}^\infty\,\frac{a_{n0}}{\omega_{n0}}
 \,\vec{u}_{n00}(\vec{x})\,g_{n0}^{(0,0)}(t)\nonumber \\[1 ex]
 & + & \sum_{n=1}^\infty\,\frac{a_{n2}}{\omega_{n2}}\,\left[\sum_{m=-2}^2
 \vec{u}_{n2m}(\vec{x})\,g_{n2}^{(2,m)}(t)\right]\;,
 \label{eq5.15}
\end{eqnarray}
where

\begin{subeqnarray}
 a_{n0} & = & -\frac{1}{M}\,\int_0^R A_{n0}(r)\,\rho\,r^3\,\D r\;,
 \label{eq5.16a}    \\*[1 em]
 a_{n2} & = & -\frac{1}{M}\,\int_0^R \left[A_{n2}(r) +
 3\,B_{n2}(r)\right]\,\rho\,r^3\,\D r\;,
 \label{eq5.16b}
\end{subeqnarray}
and

\begin{equation}
 g_{nl}^{(l,m)}(t)\equiv\int_0^t g^{(l,m)}(t')\,
 \sin\omega_{nl}(t-t')\,\D t'\;.
 \label{eq5.17}
\end{equation}

\begin{figure}[t]
\centering
\includegraphics[width=7 cm]{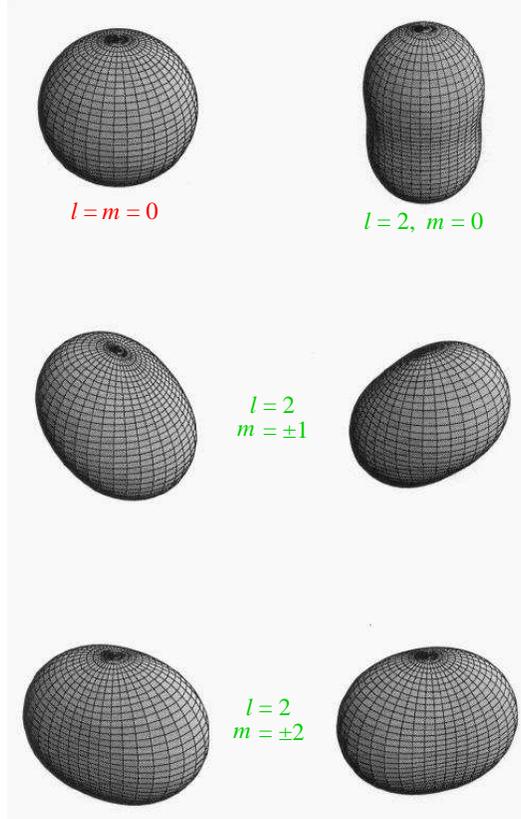}
\caption{The five quadrupole spheroidal oscillation modes of a solid elastic
sphere, and the monopole mode. Note the latter (top left) is a spherically
symmetric `breathing' mode, while the rest have non-symmetric shapes.
Because the eigenmodes~(\protect\ref{eq5.14}) are actually \emph{complex}
(the spherical harmonics for $m\neq 0$ are complex) suitable combinations of
them and their complex conjugates have been used to make plotting possible.
These \emph{shapes} are shared by \emph{all} corresponding harmonics, i.e.,
quadrupole profiles for example are those shown no matter the harmonic
number $n\/$ of their frequency $\omega_{n2}$: simply, they oscillate
faster for higher $n\/$, but always keep the represented profile.
\label{fig6}}
\end{figure}

The series expansion (\ref{eq5.15}) transparently shows that \emph{only}
monopole and qua\-dru\-pole spherical modes can possibly be excited by
an incoming GW. The monopole will of course not be excited at all if
General Relativity is the true theory of gravitation. A spherical solid
is thus seen to be the \emph{best possible} shape for a GW detector.
This is because of the optimality of the \emph{overlap coefficients}
$a_{nl}\/$ between the universal form factors $\vec{f}^{(l,m)}(\vec{x})$
and the sphere's eigenmodes $\vec{u}_{nlm}(\vec{x})$, which comes about
due to the clean \emph{multipole structure} of the latter: only the
$l\/$\,=\,0 and $l\/$\,=\,2 \emph{spheroidal} modes couple to GWs,
hence \emph{all the GW energy} is deposited into them, and only them.
Any other solid's shape, e.g.\ a cylinder, has eigenmodes most of which
have some amount of monopole/quadrupole projections in the form of the
coefficients~(\ref{eq5.8a}), and this means the incoming GW energy is
distributed amongst \emph{many modes}, thus making detection less
efficient. We shall assess quantitatively the efficiency of the
spherical detector in terms of \emph{cross section} values below.

But, as just stated, quadrupole modes are \emph{degenerate}. More
specifically, they are 5-fold degenerate, each degenerate wavefunction
corresponding to one of the five integer values $m\/$ can take between
$-2$ and $+2$. Monopole modes are instead non-degenerate. Figure~\ref{fig6}
shows the shapes of all these modes~\cite{jao} --~see the caption to
the figure for further details.

\emph{Degeneracy} is a key concept for the \emph{multimode}
capabilities of the spherical detector. For, as explicitly
shown by equation~(\ref{eq5.15}), monopole and quadrupole
\emph{detector} modes are driven by one and five \emph{GW amplitudes},
respectively, i.e., $g^{(0,0)}(t)$ and $g^{(2,m)}(t)$. Therefore, if
one could \emph{measure} the amplitudes of these modes, i.e., the
amplitudes of the deformations displayed in figure~\ref{fig6},
then a \emph{complete deconvolution} of the GW signal would be
accomplished. This is a \emph{unique} feature of the spherical
antenna, which is not shared by any other GW detector: it enables
the determination of \emph{all} the GW amplitudes, not just a
combination of them, no matter where the signal comes from.
In section~\ref{sub5.3} below I shall give a more detailed review
of how the multimode capability can be implemented in~practice.

\subsubsection{Cross sections}
\label{sub5.2a}

The general definition (\ref{eq5.11}) applies in this case, too.
Since cross section is a \emph{frequency} dependent concept, and
since quadrupole modes are degenerate, it is clear that energies
deposited in \emph{each} of the five degenerate modes of a given
frequency harmonic must be added up to obtain $E(\omega_{n2})$ for
that mode. Such energy must be calculated by means of volume
integrals --~to add up the energies of all differentials of mass
throughout the solid~--, the details of which I omit here. The
final result turns out to be a remarkable one~\cite{lobo}: cross
sections factorise in the form

\begin{equation}
 \sigma_{\rm abs}(\omega_{nl}) =
 {\cal K}_l(\aleph)\,\frac{GMv_t^2}{c^3}\,(k_{nl}a_{nl})^2
 \qquad (l=0\mbox{\ or\ } 2)\;,
 \label{eq5.18}
\end{equation}
where $GMv_t^2/c^3$ is a characteristic of the sphere's material\footnote{
$v_t\/$ is the so called `transverse speed of sound', and is related
to the true speed of sound, $v_s$, by the formula

\[
v_t = \left(2+2\sigma_P\right)^{-1/2}\,v_s\;,
\]
with $\sigma_P\/$ the material's Poisson ratio.},
and $(k_{nl}a_{nl})$ a dimensionless quantity associated with the
$\{nl\}$-th~frequency harmonic; finally, and this is the stronger
theoretical point of this expression, ${\cal K}_l(\aleph)$ is a
coefficient which is characteristic of the underlying theory of
GWs, symbolically indicated with $\aleph$. For example, if the
latter is General Relativity (GR) then

\begin{equation}
 \aleph = {\rm GR} \Rightarrow \left\{\begin{array}{l}
 {\cal K}_0(\aleph) = 0 \;, \\[0.7 em]
 {\cal K}_2(\aleph) = \mbox{\Large $\frac{16\,\pi^2}{15}$}\;,
 \end{array}\right.
 \label{eq5.19}
\end{equation}
while if it is e.g.\ Brans-Dicke~\cite{bd61} then these expressions
get slightly more complicated~\cite{bbc,svino},~etc.

\begin{table}[t]
\caption{Compared characteristics and cross sections for a cylindrical
and a spherical GW detector of like fundamental frequencies. Note that
the cylinder is assumed to be optimally oriented, i.e., with its axis
perpendicular to the GW incidence direction.
\label{tab1}}
\vspace{1 mm}
\begin{center}
\begin{tabular}{cc}
\hline \\[-1 em]
{\sf Cylinder} & {\sf Sphere} \\[0.7 ex] \hline \\
$\nu_{1}$ = 910 Hz & $ \left\{\begin{array}{l}
			\nu_{12} = 910 \ {\rm Hz} \\
			\nu_{22} = 1747 \ {\rm Hz}
			\end{array}\right. $ \\[1 em]
$ \left\{\begin{array}{l}
   L = 3.0 \ {\rm metres} \\
   D = 0.6 \ {\rm metres}
  \end{array}\right. $  &  $2R\/$ = 3.1 metres \\[1.6 em]
$M_c$ = 2.3 tons & $M_s$ = 42 tons    \\[1 em]
\ \ \ \ $\sigma_{1} = 4.3$$\times$$10^{-21}$ cm$^2$ Hz\ \ \ \  &
\ \ \ \ $\left\{\begin{array}{l}
  \sigma_{12} = 9.2\!\times\!10^{-20}\;{\rm cm}^2\ {\rm Hz} \\
  \sigma_{22} = 3.5\!\times\!10^{-20}\;{\rm cm}^2\ {\rm Hz}
 \end{array}\right.$\ \ \ \  \\[1.6 em]
({\bf Optimum orientation})  &  ({\bf Omnidirectional}) \\[0.6 ex] \hline
\end{tabular}
\end{center}
\end{table}

Sticking to GR, a few illustrative figures are in order. They are
shown in table~\ref{tab1}, where a material of aluminum Al5056 alloy
has been chosen. It appears that a sphere having the same fundamental
frequency ($\nu_{12}$) as a cylinder ($\nu_1$) is about 20 times more
massive, and this results in a significant improvement in cross
section, since it is proportional to the detector mass. A spherical
detector is therefore almost \emph{one order of magnitude} more sensitive
than a cylinder in the same frequency band --~obviously apart from
the important fact that the sphere has \emph{isotropic} sensitivity.

\begin{figure}[t]
\centering
\includegraphics[width=9 cm]{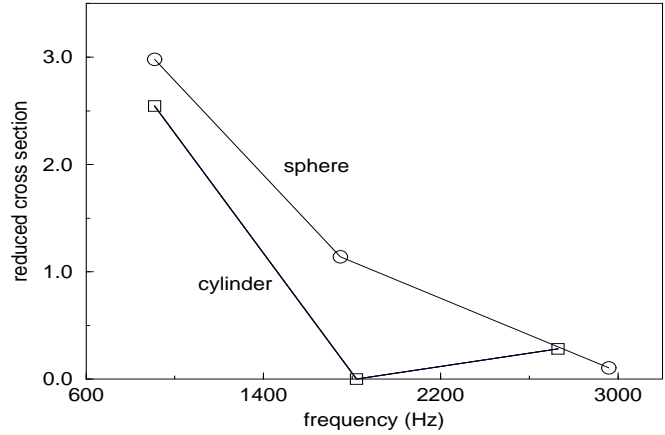}
\caption{Cross section \emph{per unit mass} of a cylinder and a sphere
of like fundamental frequencies, in units of $Gv_s^2/c^3$. Note the
(slightly) better figure (17\,\%) for the lowest mode in the sphere,
as well as the appreciable value in the second harmonic of the latter,
in sharp contrast with the \emph{null} coupling of this mode in the
cylinder. Third harmonics show a considerable reduction in sensitivity.
\label{fig7}}
\end{figure}

But there is more to this. Table~\ref{tab1} also refers to the sphere's
cross section in its \emph{second} higher quadrupole harmonic frequency,
$\nu_{22}$ --~almost twice the value of the first, $\nu_{12}$. It is
very interesting that cross section at this second frequency is
\emph{only} 2.61 times smaller than that at the first~\cite{clo} while,
as stressed in section~\ref{sub2.1}, it is \emph{zero} for the second
mode of the cylinder. Figure \ref{fig7} shows a plot of the cross sections
\emph{per unit mass} of a cylinder and a sphere of like fundamental
frequencies. It graphically displays the numbers given in the table,
but also shows that, even per unit mass, the sphere is a better detector
than a cylinder --~its cross section `curve' stays above the cylinder's.
In particular, the \emph{first} quadrupole resonance turns out to
have a cross section which is 1.17 times that of the cylinder (per
unit mass, let me stress again)~\cite{clo}, i.e., $\sim$\,17\,\% better.
This constitutes the \emph{quantitative} assessment of the discussion
in the paragraph immediately following equation~(\ref{eq5.17}).

\subsection{The \bfit{motion sensing} problem}
\label{sub5.3}

In order to determine the actual GW induced motions of an elastic
solid a \emph{motion sensing} system must be set up. In the case of
currently operating cylinders this is done by what is technically
known as \emph{resonant transducer}~\cite{paik}. The idea of such
device is to couple the \emph{large} oscillating cylinder mass (a few
tons) to a \emph{small} resonator (less than 1 kg) whose characteristic
frequency is accurately tuned to that of the cylinder. The \emph{joint}
dynamics of the resulting system $\{$cylinder + resonator$\}$ is a
two-mode \emph{beat} of nearby frequencies given by

\begin{equation}
 \omega_\pm\simeq\omega_0\left(1\pm\frac 12\,\eta^{1/2}\right)\;,
 \label{eq5.20}
\end{equation}
where $\omega_0$ is the frequency of either oscillator when uncoupled
to the other, and $\eta\equiv M_{\rm resonator}/M_{\rm cylinder}$. The
key concept of this device is the \emph{resonant energy transfer},
which flows back and forth between cylinder and sensor with the period
of the beat, i.e., $2\pi(\eta\omega_0)^{-1/2}$. This means that, because
the sensor's mass is very small compared to that of the cylinder, the
amplitude of its oscillations is enhanced by a factor of $\eta^{-1/2}$
relative to those of the cylinder, whence a \emph{mechanical amplification}
factor is obtained \emph{before} the sensor oscillations are converted to
electrical signals, and further processed --~see a more detailed account
of these principles in~\cite{amaldi2}.

The same principles can certainly be applied to make resonant motion
sensors in a spherical antenna. In this case, however, a special bonus is
there, associated to the \emph{degeneracy} of the quadrupole frequencies:
because all five quadrupole modes oscillate with the \emph{same} frequency,
it is possible to attach five (or more) identical resonators, tuned to
a given quadrupole frequency, at suitable positions on the sphere surface,
thus taking \emph{multiple samples} of the sphere's motion. This makes
possible to retrieve the oscillation amplitudes of the five degenerate
modes --~figure~\ref{fig6}~--, and thereby of the GW quadrupole
amplitudes $g^{(l,m)}(t)$, since both are linearly related through
equation~(\ref{eq5.15}).

A \emph{single} spherical antenna can thus deconvolve completely the
quadrupole GW signal, and do so with \emph{isotropic sky coverage}.
These characteristics are \emph{unique} to the spherical detector,
and they make it a \emph{theoretically} superior system compared
to either interferometers or Weber bars. In addition, a sphere can
naturally measure the amplitude of the non-degenerate monopole mode,
as it is conceptually simple to sense the amplitude of an isotropically
breathing pattern.

\begin{figure}[t]
\centering
\includegraphics[width=4 cm]{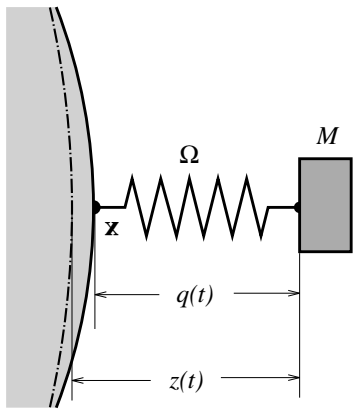}
\caption{Schematic diagramme of the coupling between a solid sphere and
a resonator, modeled as a small mass linked to a spring attached to the
sphere's surface. The dashed-dotted arc line on the left indicates the
position of the \emph{undeformed} sphere's surface, and the solid arc
its \emph{actual} position.
\label{fig8}}
\end{figure}

The conceptual idea of a resonant sensor is shown in figure~\ref{fig8},
and the equations of motion for such a system are~\cite{mnras}

\begin{subeqnarray}
 \varrho \frac{\partial^2 \vec{u}}{\partial t^2} & = & \mu\nabla^2 \vec{u}
 + (\lambda+\mu)\,\nabla(\nabla{\bf\cdot}\vec{u}) +
 \nonumber \\
 & & \sum_{b=1}^J M_b\Omega_b^2\,\left[z_b(t)-u_b(t)\right]\,
 \delta^{(3)}(\vec{x}-\vec{x}_b)\,\vec{n}_b
 + \vec{f}_{\rm GW}(\vec{x},t)\;,
 \label{eq5.20a}  \\*[0.7 em]
 \ddot{z}_a(t) & = & -\Omega_a^2\,
 \left[z_a(t)-u_a(t)\right] + \xi_a^{\rm GW}(t)\ ,
 \qquad a=1,\ldots,J\;,
 \label{eq5.20b}
\end{subeqnarray}
where $M_a$ and $\Omega_a$ are the mass and characteristic frequency
of the $a$-th resonator, $\vec{f}_{\rm GW}(\vec{x},t)$ is the GW tide
on the sphere -- see~(\ref{eq5.6}) --, and $\xi_a^{\rm GW}(t)$ is the
GW induced tidal acceleration on the resonator itself, relative to the
centre of the sphere; $\delta^{(3)}$ is the three dimensional Dirac
density, i.e., point-like connections between sphere and sensors are
assumed. The mathematical detail of the analysis of these equations
is somewhat sophisticated. The interested reader will find complete
information in~\cite{mnras}; the rest of this section will be devoted
to a brief discussion of the main conclusions of that analysis.

First thing to stress is that equations~(\ref{eq5.20a}-b) cannot be
solved analytically, they must instead be solved by a \emph{perturbative}
procedure. The small perturbation parameters are the ratios

\begin{equation}
 \eta_a\equiv\frac{M_a}{M}\ ,\qquad\eta_a\ll 1\ ,\qquad a=1,\ldots,J\;,
 \label{eq5.21}
\end{equation}
where $M\/$ is the sphere's total mass. Actually, the analysis assumes
that the resonators are all \emph{identical}, any deviations from this
being eventually assessed by suitable methods~\cite{mnras,jm97}. The
fundamental result links the spring deformations to the the GW amplitudes
$g^{(l,m)}(t)$ by the following formula, expressed in terms of
\emph{Laplace transforms} -- noted with a caret (\,$\hat{}$\,):

\begin{equation}
 \hat q_a(s) = \eta^{-1/2}\,\sum_{l,m}\,\hat\Lambda_a^{(lm)}(s;\Omega)\,
 \hat g^{(l,m)}(s)\ ,\qquad a=1,\ldots,J\;,
 \label{eq5.22}
\end{equation}
where it is assumed that the resonators' frequency $\Omega$ is tuned
to either a mono\-pole or a quadrupole harmonic of the sphere. The
\emph{transfer functions} $\hat\Lambda_a^{(lm)}(s;\Omega)$ naturally
depend on whether a monopole or a quadrupole mode is selected for
resonator tuning; I will quote here only the quadrupole case, as it is
the most interesting one -- see again full information in~\cite{mnras}:

\begin{eqnarray}
 \hat\Lambda_a^{(lm)}(s;\omega_{n2}) & = & (-1)^J\,\sqrt{\frac{4\pi}{5}}
 \;a_{n2}\,\sum_{b=1}^J\,\left\{\sum_{\zeta_c\neq 0}\,\frac{1}{2}\left[
 \left(s^2+\omega_{c+}^2\right)^{-1} - \left(s^2+\omega_{c-}^2\right)^{-1}
 \right]\right.\nonumber \\
 & \times & \left.\frac{v_a^{(c)}v_b^{(c)*}}{\zeta_c}\right\}\,
 Y_{2m}(\vec{n}_b)\,\delta_{l2}\;,
 \label{eq5.23}
\end{eqnarray}
where $v_a^{(c)}$ is the $c\/$-th normalised eigenvector of the matrix
$P_2(\vec{n}_a\!\cdot\vec{n}_b)$, associated to its \emph{non-zero}
eigenvalue $\zeta_c^2$, $P_2$ is a Legendre polynomial, and $\vec{n}_a$
is the position of the $a\/$-th resonator on the sphere's surface.
Finally,

\begin{equation}
 \omega_{a\pm}^2 = \omega_{n2}^2\,\left(1\pm\sqrt{\frac{5}{4\pi}}\,
 \left|A_{n2}(R)\right|\,\zeta_a\,\eta^{1/2}\right)\ ,
 \qquad a=1,\ldots,J\;.
 \label{eq5.24}
\end{equation}

Equations (\ref{eq5.22})--(\ref{eq5.24}) are the key to the GW signal
deconvolution problem: they show that \emph{beats} occur around the
tuned frequency ($\omega_{n2}$ in this case), and that the resonators
oscillate with amplitudes enhanced by a factor $\eta^{-1/2}$, as indeed
expected. Note that these beats have frequencies which depend on the
resonators' positions $\vec{n}_a$, as shown by the presence of
the~$\zeta_a$ coefficient in~(\ref{eq5.24}).

\begin{figure}[b]
\centering
\includegraphics[width=9 cm]{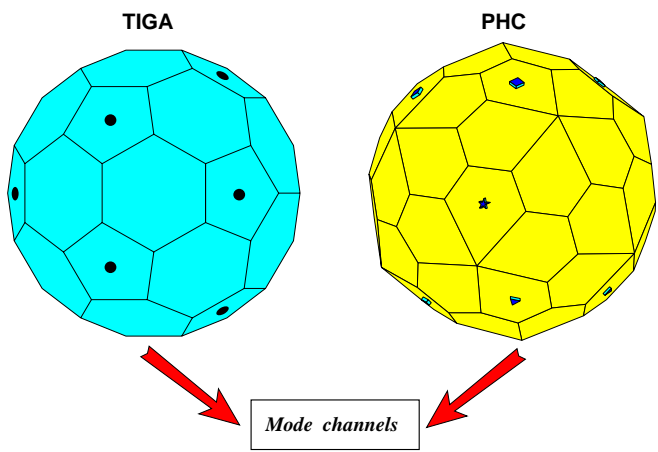}
\caption{The {\sl TIGA\/} and {\sl PHC\/} resonator distributions. In
the former, \emph{six} sensors are attached to the pentagonal faces of
a truncated icosahedron, while in the latter there are \emph{two} sets
of \emph{five} quadrupole sensors, respectively tuned to the first
quadrupole frequency (squares) and to the second (triangles); there
is an 11-th sensor (star) which is tuned to a monopole frequency. The
relevant common characteristic of these layouts is that they enable
the definition of \emph{mode channels} --~see text for details.
\label{fig9}}
\end{figure}

The deconvolution problem consists in inferring the GW amplitudes
$\hat g^{(l,m)}(s)$ from the telescope's readouts $\hat q_a(s)$.
Thus \emph{at least} 5 sensors must be attached to the sphere's
surface if e.g.\ the 5 quadrupole amplitudes are looked for: once
the corresponding five $\hat q_a(s)$ are \emph{measured} the
system~(\ref{eq5.22}) is solved for $\hat g^{(2,m)}(s)$, that's it.

Crucial at this point is to decide \emph{where} to implant the
resonators, as such decision bears fundamentally on the simplicity,
or even the possibility of solving the posed problem. There are two
major proposals in the literature for this, and they are displayed
in figure~\ref{fig9}. They permit the definition of so called
\emph{mode channels}, which are linear combinations of the system
readouts $\hat q_a(s)$ which are directly proportional to the GW
amplitudes $\hat g^{(2,m)}(s)$. They happen to be of the form~\cite{mnras}

\begin{equation}
 \hat y^{(m)}(s) = \sum_{a=1}^{5\ {\rm or}\ 6}\,v_a^{(m)*}\hat q_a(s)\ ,
 \qquad m=-2,\ldots,2\;,
 \label{eq5.25}
\end{equation}
both for {\sl TIGA\/} and {\sl PHC}. The actual result of these linear
combinations is the following:

\begin{equation}
 \hat y^{(m)}(s) = \eta^{-1/2}\,a_{n2}\,
 \frac{1}{2}\left[\left(s^2+\omega_{m+}^2\right)^{-1} -
 \left(s^2+\omega_{m-}^2\right)^{-1}\right]\,\hat g^{(2,m)}(s)\;,
 \label{eq5.26}
\end{equation}
i.e., they are \emph{convolution products} of the signal and the system
beats.

This formula appears to be very powerful, as it shows that suitable
sensor systems enable a \emph{single} spherical detector to fully
deconvolve \emph{all} the GW amplitudes. One should however be careful
about this conclusion, for the formula \emph{also} indicates that the
relevant information \emph{can only be obtained at the resonance
frequencies} $\omega_{m\pm}$ --~in an ideal, non-dissipative system.
In a real system, resonant linewidths are never infinitely sharp,
they have instead a certain breadth; this actually makes possible
the observation across wider bandwidths, provided the amplifier noise
can be kept sufficiently small --~I shall briefly come to this below
in section~\ref{sec.6}.

\subsection{\bfit{Hollow} and \bfit{dual} spheres}
\label{sub5.4}

The real merit of the just described spherical GW detector comes from
its \emph{symmetry}. A \emph{hollow} sphere does of course share the
symmetry properties of a solid sphere, and one might therefore expect
it to be an interesting alternative, too. A detailed study of the
performance of a hollow spherical GW detector can be found in~\cite{vega}.
The added bonus of a hollow sphere is that there is one more structural
parameter one can adjust to enhance this or that property, and this is
of course the \emph{thickness} of the spherical shell. It appears that
cross sections at e.g.\ different quadrupole harmonics have characteristic
behaviours when plotted as functions of thickness; also, one can decide
to attach the motion sensors to the inner \emph{or} to the outer side of
the shell, as their GW induced oscillations have different amplitudes.

\begin{figure}[t]
\centering
\includegraphics[width=5 cm]{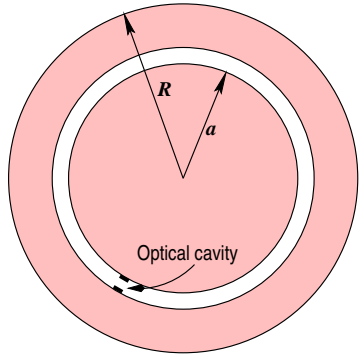}
\caption{Conceptual scheme of the \emph{dual sphere}: the GW signal
drives the facing surfaces between the nested spheres into oscillatory
motions; when the signal frequency falls between the resonances of
the spheres, oscillations happen with phase opposition, thereby
enhancing the device's response. Motion sensing can be accomplished
e.g.\ by non-resonant optical transducers.
\label{fig10}}
\end{figure}

To actually build and suspend a hollow sphere in the laboratory may
be a difficult task from a technological point of view. Recently,
though, a new concept spherical detector appeared in the
literature~\cite{dual}: this is called \emph{dual sphere}, and
consists in a solid sphere nested inside a hollow one, concentric
with it and with a narrow gap between them --~figure~\ref{fig10}.
An incoming GW drives \emph{both} spheres into oscillation. Clearly,
the inner solid sphere has a higher first (quadrupole) mode frequency 
than the outer hollow one, therefore an incoming GW with a frequency
\emph{between} those two will drive the oscillations of the spheres
with opposite phase\footnote{
As shown in textbooks on Mechanics, see e.g.~\cite{symon}, there is
phase change in an oscillator's response to a periodic excitation
as the frequency of the latter shifts from below the oscillator's
natural frequency ($\Omega$, say) to above it; the transition region
has a width of order $\Omega/Q\/$ centred at $\Omega$, where $Q\/$
is the oscillator's mechanical quality factor, about 10$^7$ or more
in GW detectors.}, thus enhancing the signal by a rough factor of~2.

The motion sensing in either hollow or dual spheres is conceptually
analogous to that in the solid sphere, with the added flexibility that
in the hollow piece one can choose to sense the displacements of either
side of the shell. Actually, though, it appears that non-resonant
detectors appear constitute a better choice in dual spheres, for
this enables a significant bandwidth enlargement~\cite{dual}.

Let me briefly discuss now, for completeness, a few essentials of GW
detector sensitivity.

\section{GW detector sensitivities}
\label{sec.6}

So far I have only discussed the \emph{theoretical} basis of the workings
of GW detectors, whether interferometric or acoustic, yet have made
no mention of the \emph{practical} difficulties of getting them
actually working\ldots

The extreme weakness of any expected GW signals arriving in the
Earth~\cite{thorne} is in fact a source of such truly difficult problems
that it has prevented GW detectors from sighting a real signal in the
last 40 years, since the times of J.~Weber. Local \emph{detector noise}
has to date overwhelmed any signals possibly hitting the antennas, and
therefore the technological challenge has been for years, and still is
today, to reduce that noise to the level where it can be filtered out
with a meaningful signal to noise ratio~\cite{klm93}.

During the last decade or so, a number of people in different countries
worldwide have managed to get important GW detection research projects
funded which constitute a major step forward in detector technology.
Their goal is to improve the sensitivity to the point where a
significant \emph{event rate} becomes available to the GW astronomer.
We thus find such laboratories as {\sl VIRGO\/} (a French-Italian
collaboration), {\sl LIGO\/} (USA alone), or {\sl LISA\/} (a space
mission, jointly funded by {\sl NASA\/} and {\sl ESA}, the European
Space Agency). In addition to those, somewhat smaller experiments
are {\sl GEO-600\/} (a German-British venture) and {\sl TAMA\/}
(the Japanese project, currently making strong progress in both
and stability).

So much for \emph{interferometric} antennas. But endeavours have not
declined in the \emph{acoustic detector} arena, either. In fact, the five
acoustic detectors of the Weber type ({\sl NAUTILUS, EXPLORER, AURIGA,
ALLEGRO and NIOBE\/}) constitute the \emph{only working systems} in
the world today. Unfortunately, though, they are only sensitive to
catastrophic events happening in our galaxy, with a far too low
occurrence rate. These detector systems are periodically upgraded,
and their sensitivity has gone up a few orders of magnitude since
their origins. As already stressed in section~\ref{sec.1}, new
generation \emph{spherical} GW detectors are being programmed in
Brazil, Holland and Italy, and these should suddenly improve over
bars by at least one more order of magnitude.

\begin{figure}[b]
\centering
\includegraphics[width=10 cm]{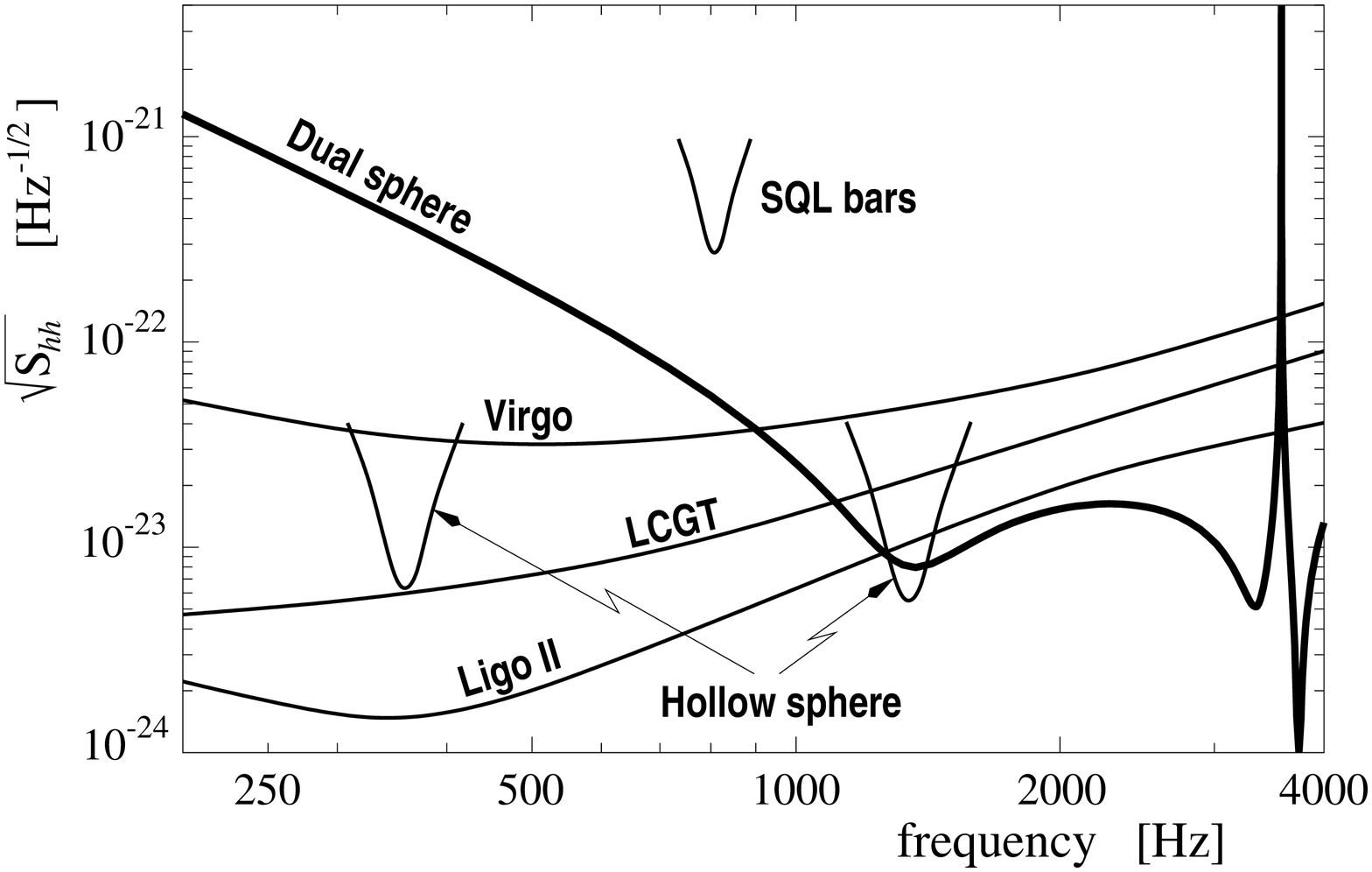}
\caption{Spectral strain sensitivities of various future generation
GW detectors, including the ``dual sphere''.
\label{fig.11}}
\end{figure}

Figure~\ref{fig.11} contains a recent plot of GW sensitivity of
various earth based detectors, therefore in the frequency range
near 1\,kHz. On the other hand, {\sl LISA\/} will be sensitive at
frequencies far away from the range plotted in figure~\ref{fig.11}
%more specifically the milli-Hz range
--~see {\sl LISA}'s web site at {\tt http://sci.esa.int/home/lisa}
%
%With the exception of the dual sphere~\cite{dual},
%bibliographic reference to GW detectors is not provided, as the
%reader may advantageously visit the web sites of the different
%projects, where he/she will find updated information.
%
and figure~\ref{fig.12} below. What we see in ordinates in plots like
the one in figure~\ref{fig.11}, which are standard in GW science, is
a \emph{spectral density} --~or, rather, its \emph{square root}. It
is to be understood as follows.

\emph{Noise} of whatever origin causes any detector to generate random
outputs --~\emph{stochastic} time series, as this is technically known.
This noise competes with any GW signals which may be present in the
antenna readout, obscuring their detection. Noise of course does not
come from the sky, it rather gets added to the GW signal at almost all
the different stages of the detection process. For example, the GW
induced oscillations of a given mass compete with thermal oscillations
of that mass and its suspension systems; then there is noise in the
conversion from mechanical or optical GW signals to electrical signals;
then there is noise introduced by the electronics in the amplifiers of
the latter; and so forth\ldots

In the end, the detector output is a certain physical magnitude (e.g.\ 
a voltage) which consists of a number of various superimposed sources
of noise plus, possibly, a GW signal \emph{converted} to volts by the
detector hardware. One infers the value of the \emph{actual} dimensionless
GW amplitude $h(t)$, say, from the output voltage if one knows the
precise physics of the transduction process --~and this is obviously
the case in any useful instrument.

Conversely, it is also possible to \emph{back-convert} all the sources
of noise voltage picked up across the various detector stages to an
equivalent dimensionless ``GW noise'', which gets directly added to
real GW signals, and travels through an ideally noiseless detector.
This artifact is expedient because, for a fixed antenna, it enables
a quick assessment of the detectability of a given GW signal, normally
calculated by methods of gravitation theory, by direct comparison with
suitably constructed detector characteristic curves -- such as those
in figure~\ref{fig.11}.

Let us then call $x(t)$ any one of the detector's readout channels,
\emph{back-con\-verted} to a GW amplitude by the above described procedure.
We split this up into a GW signal proper $h(t)$ plus a noise term $n(t)$:

\begin{equation}
 x(t) = h(t) + n(t)\;.
 \label{eq5.27}
\end{equation}

For stationary Gaussian noise, the statistical properties of $n(t)$
are encoded in its \emph{spectral density} function, $S_h(\omega)$,
which is the Fourier transform of the \emph{autocorrelation function}

\begin{equation}
 R(\tau)\equiv\langle x(t)\,x(t+\tau)\rangle\ ,\qquad
 S_h(\omega)\equiv\int_{-\infty}^\infty\,R(\tau)\,\E^{\I\omega\tau}\,\D\tau\;,
 \label{eq5.28}
\end{equation}
where $\langle -\rangle$ stands for \emph{ensemble average}~\cite{kay}.
It can be shown~\cite{klm93} that the \emph{optimum filter} to extract
the signal $h(t)$ from the system readout $x(t)$ is the so called
\emph{matched filter}, whose transfer function is the signal's
Fourier transform $\tilde h(\omega)$ divided by the spectral density
$S_h(\omega)$, and the \emph{detection threshold} can be set from
the integrated signal to noise ratio:

\begin{equation}
 SNR = \int_{-\infty}^\infty\,\frac{|\tilde h(\omega)|^2}{S_h(\omega)}\,
 \frac{\D\omega}{2\pi}\;.
 \label{eq5.29}
\end{equation}

The GW signal $h(t)$ is a dimensionless quantity, as it measures a
perturbation of the Minkowski metric, see equation~(\ref{eq2.2}).
Therefore the spectral density $S_h(\omega)$ has dimensions of time, or
inverse frequency, Hz$^{-1}$, according to the definitions~(\ref{eq5.28}).
Now, signal to noise ratio as defined by the integral~(\ref{eq5.29}) is
made up of the contributions of the ratio between the signal power
$|\tilde h(\omega)|^2$ to the noise power $S_h(\omega)$ at all frequencies;
the idea of the graphical representation in figure~\ref{fig.11} is thus
to show which is the level of noise at different frequencies by means of
an rms quantity, such as is the square root of the noise spectral density.
The appropriate units for this representation are accordingly Hz$^{-1/2}$,
as indicated.

\begin{figure}[t]
\centering
\includegraphics[width=12 cm]{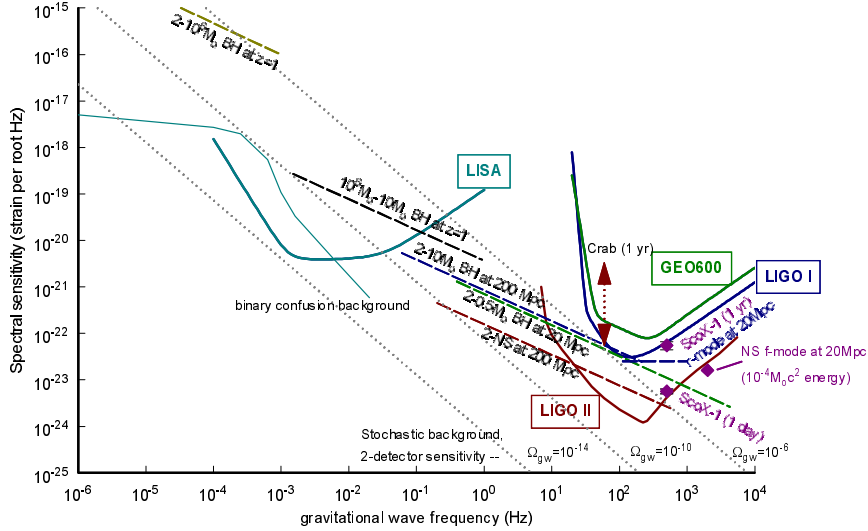}
\caption{Root mean square noise spectral densities, referred to GW 
amplitudes, for some of the upcoming interferometric GW detectors,
together with the spectral intensities of various signals. The
latter are estimated by numerical calculation, while the noise is
modeled on the basis of its origin and instrumental characteristics.
See~\protect\cite{bfs}.
\label{fig.12}}
\end{figure}

Different interesting sources of GWs are being considered by other
authors in this volume, so I will not go into such matters here. It
is nevertheless instructive to present an example graph of a few
signals on top of various detectors' sensitivity curves in order to
get a picture of the actual possibilities of each instrument, and
also to grasp what are the spectral orders of magnitude of different
GW signals. One such plot is presented in figure~\ref{fig.12}. This
is, let me insist, the standard way to assess the detectability of
the different GW sources.

\section{Concluding remarks}
\label{sec.7}

This article is a brief review on the nature of the interactions of GWs
with \emph{test} particles and \emph{test} electromagnetic fields, as
they specifically happen in currently conceived detection devices. While
the fundamental principles are not new in themselves, their application
in actual systems is still in many cases subject of research, as we
have seen. A thorough understanding of these matters is absolutely
essential for an adequate interpretation of the antenna readouts,
the more so if one considers the extreme \emph{weakness} of any
signals reaching us from even the most intense sources.

I have omitted any detailed discussion of the practical problems faced
by real detector building. This is a major research field in itself,
of an intrinsically sophisticated and multidisciplinary nature
-- involving such fundamental issues as quantum measurement limits
and techniques, or Quantum Optics~\cite{brag}. But it is of course
not directly related to Astrophysics or Relativity\ldots\ I have
however considered appropriate to summarily brief the interested
reader on how detector sensitivities are defined, and on which are
the detection thresholds in current state of the art GW detectors:
this is a key issue for an astrophysicist/relativist wishing to
assess the detectability of a given GW signal, whether by
existing instruments or future planned.

GW detection endeavours have been the subject of intensive research
during the last 4 decades, though the ultimate goal of sighting GW
events has still not been accomplished. While this may look like a major
failure, one may not forget that detector sensitivities have gone up
by a remarkable \emph{six orders of magnitude} (in energy, three in GW
amplitude) since the very first telescopes constructed by J.~Weber.
Also, a look at the trend in this development indicates that we are
getting closer to the obejctive.

Prospects look now better than ever yet the real challenge is still there\ldots


\begin{thebibliography}{99}
\addcontentsline{toc}{section}{References}

\bibitem{ein18} A.\ Einstein: Sitz.\ Ber.\ K\"on.\ Preus.\ Ak.\ Wiss.,
	p.\ 688 (1916), and p.\ 154 (1918)

\bibitem{mtw} C.W.\ Misner, K.S.\ Thorne, J.A.\ Wheeler: \emph{Gravitation}
	(Freeman, San Francisco~1973)

\bibitem{will} C.M.\ Will: \emph{Theory and experiment in Gravitational
	Physics} (Cambridge University Press, Cambridge 1993)

\bibitem{weber} J.\ Weber: Phys.\ Rev.\ Lett.\ {\bf 22}, 1320 (1969);
	{\bf 24}, 276 (1970); {\bf 25}, 180 (1970)

\bibitem{jweb2} J.\ Weber: `Supernova 1987\,A Rome and Maryland GW
	radiation antenna observations'. In: \emph{Gravitational wave
	experiments}, Proceedings of the 1st Edoardo Amaldi Conference,
	ed.\ by E.\ Coccia, G.\ Pizzella, F.\ Ronga (World Scientific,
	Singapore 1995), p.\ 416

\bibitem{gh71} G.W.\ Gibbons, S.W.\ Hawking: Phys.\ Rev.\ D {\bf 4},
	2191 (1971)

\bibitem{weiss} R.\ Weiss: Quarterly Progress Report, Research Laboratory
	of Electronics {\bf 105}, 54 (MIT 1972)

\bibitem{ron} R.\ Drever: `Interferometric detectors for gravitational
	radiation'. In: \emph{Gravitational radiation}, ed.\ by N.\ 
	Deruelle and T.\ Piran (North Holland, Amsterdam 1983), p.~321

\bibitem{igec} G.A.\ Prodi et al.: Int.\ J.\ Mod.\ Phys.\ {\bf D9},
	237 (2000)

\bibitem{msch} O.D.\ Aguiar \emph{et al.}: `The first phase of
	the Brazilian Graviton project'. In: \emph{Gravitational waves},
	Proceedings of the 3rd Edoardo Amaldi Conference, ed.\ by S.\ 
	Meshkov (World Scientific, Singapore 2000), p.\ 413

\bibitem{mgrail} G.\ Frossati: `{\sl MINIGRAIL}, a sensitive spherical
	gravitational wave antenna for frequencies around 3.6 kHz',
	Projectruimte Normale Programma FWO (Leiden~2000)

\bibitem{sfera} P.\ Astone \emph{et al.}: `{\sl SFERA\/}: Proposal for
        a spherical GW detector' (Roma 1997)

\bibitem{ht75} R.A.\ Hulse, J.H.\ Taylor: Astrophys.\ Jour.\ {\bf 195},
	L51 (1975)

%\bibitem{wt89} J.H.\ Taylor \& J.M.\ Weisberg: Astrophys.\ Jour.
%	{\bf 345}, 434 (1989)

\bibitem{tay94} J.H.\ Taylor: Rev.\ Mod.\ Phys.\ {\bf 66}, 711 (1994)

%\bibitem{f3m} T.A.\ Apostolatos et al.: Phys.\ Rev.\ Lett.\ {\bf 70}
%	2984 (1993)

%\bibitem{cofo} E.\ Coccia, V.\ Fafone: Phys.\ Lett.\ A {\bf 213}, 16 (1996)

\bibitem{wein} S.\ Weinberg: \emph{Gravitation and Cosmology\/}
	(Wiley \& Sons, New York 1972)

\bibitem{grif} J.B.\ Griffiths: \emph{Colliding plane waves in General
	Relativity} (Clarendon Press, Oxford 1991)

\bibitem{bd61} C.\ Brans, R.H.\ Dicke: Phys.\ Rev.\ {\bf 124}, 925 (1961)

\bibitem{gazta} E.\ Gazta\~naga, J.A.\ Lobo: Astroph.\ Jour.\ {\bf 548},
	47 (2001)

\bibitem{el73} D.M.\ Eardley, D.L.\ Lee, A.P.\ Lightman: Phys.\ Rev.\ D
	{\bf 8}, 3308 (1973)

\bibitem{chan} S.\ Chandrasekhar: \emph{The mathematical theory of
	Black Holes} (Pergamon Press, Oxford 1983)

\bibitem{brian} B.J.\ Meers: Phys.\ Rev.\ D {\bf 38}, 2317 (1988)

\bibitem{cqg} J.A.\ Lobo: Class.\ Quan.\ Grav.\ {\bf 9}, 1385 (1992)

\bibitem{fara} F.I.\ Cooperstock, V.\ Faraoni: Class.\ Quan.\ Grav.\
	{\bf 10}, 1189 (1993)

\bibitem{pizz} G.\ Pizzella: \emph{Fisica Sperimentale del Campo
	Gravitazionale} (Nuova Italia Scientifica, Roma 1993)

\bibitem{ll70} L.D.\ Landau, E.M.\ Lifshitz: {\it Theory of Elasticity}
	(Pergamon Press, Oxford 1970)

\bibitem{lobo} J.A.\ Lobo: Phys.\ Rev.\ D {\bf 52}, 591 (1995)

\bibitem{ric} R.G.\ Hier, S.N.\ Rasband: Astroph.\ Jour.\ {\bf 195},
	507 (1975).

\bibitem{Ras} S.N.\ Rasband: Jour.\ Acous.\ Soc.\ Am.\ {\bf 57}, 899 (1975)

\bibitem{expl} P.\ Astone {\it et al.}: Phys.\ Rev.\ D {\bf 47}, 362 (1993)

\bibitem{naut} P.\ Astone {\it et al.}: Astropart.\ Phys.\ {\bf 7}, 231 (1997)

\bibitem{st} S.V.\ Dhurandhar, M.\ Tinto: Mon.\ Not.\ Roy.\ Astr.\ Soc.\ 
	{\bf 234}, 663 (1988), and {\bf 236}, 621 (1989)

\bibitem{bars} Z.A.\ Allen {\it et al.}: Phys.\ Rev.\ Lett.\ {\bf 85},
	5046 (2000)

\bibitem{fo71} R.\ Forward: Gen.\ Rel.\ Grav.\ {\bf 2}, 149 (1971)

\bibitem{ad75} N.\ Ashby, J.\ Dreitlein: Phys.\ Rev.\ {\bf D12}, 336 (1975).

\bibitem{wp77} R.V.\ Wagoner, H.J.\ Paik: `Multimode detection
	of gravitational waves by a sphere'. In: \emph{Experimental
        Gravitation}, Proceedings of the Pavia International Symposium,
        Accad.\ Naz.\ dei Lincei (Roma 1977)

\bibitem{jm93} W.W.\ Johnson, S.M.\ Merkowitz: Phys.\ Rev.\ Lett.\ 
	{\bf 70}, 2367 (1993)

\bibitem{phd} S.\ M.\ Merkowitz: Truncated Icosahedral Gravitational Wave
	Antenna. PhD Thesis Memoir, Louisiana State University (Baton
	Rouge 1995)

\bibitem{eug} E.\ Coccia: private communication

\bibitem{bian} M.\ Bianchi, E.\ Coccia, C.N.\ Colacino, V.\ Fafone,
        F.\ Fucito: Class.\ Quan.\ Grav.\ {\bf 13}, 2865 (1996)

\bibitem{ed60} A.R.\ Edmonds: \emph{Angular Momentum in Quantum Mechanics}
	(Princeton University Press, Princeton 1960)

\bibitem{jao} J.A.\ Ortega: Spherical gravitational wave detectors.
	PhD Thesis Memoir, University of Barcelona (Barcelona 1997)

\bibitem{bbc} M.\ Bianchi, M.\ Brunetti, E.\ Coccia, F.\ Fucito,
	J.A.\ Lobo: Phys.\ Rev.\ D {\bf 57}, 4525~(1998)

\bibitem{svino} E.\ Coccia, F.\ Fucito, J.A.\ Lobo, M.\ Salvino:
	Phys.\ Rev.\ D {\bf 62}, 044019-1 (2000)

\bibitem{clo} E.\ Coccia, J.A.\ Lobo, J.A.\ Ortega: Phys.\ Rev.\ D
	{\bf 52}, 3735 (1995)

\bibitem{paik} H.J.\ Paik: `Electromechanical transducers and bandwidth
	of resonant mass GW detectors'. In: \emph{Gravitational wave
	experiments}, Proceedings of the 1st Edoardo Amaldi Conference,
	ed.\ by E.\ Coccia, G.\ Pizzella, F.\ Ronga (World Scientific,
	Singapore 1995), p.\ 201

\bibitem{amaldi2} J.A.\ Lobo: `Spherical gravitational wave detectors and
	geometry'. In: \emph{Gravitational waves}, Proceedings of the 2nd
	Edoardo Amaldi Conference, ed.\ by E.\ Coccia, G.\ Pizzella, G.\ 
	Veneziano (World Scientific, Singapore 1998), p.\ 168

\bibitem{mnras} J.A.\ Lobo: Mon.\ Not.\ Roy.\ Astr.\ Soc.\ {\bf 316},
	173 (2000). See also {\tt gr-qc/0006055}

\bibitem{jm97} S.M.\ Merkowitz, W.W.\ Johnson: Phys.\ Rev.\ D {\bf 56},
        7513 (1997)

\bibitem{vega} E.\ Coccia, V.\ Fafone, G.\ Frossati, J.A.\ Lobo,
	J.A.\ Ortega: Phys.\ Rev.\ D {\bf 57}, 2051~(1998)

\bibitem{dual} M.\ Cerdonio, L.\ Conti, J.A.\ Lobo, A.\ Ortolan,
	L.\ Taffarello, J.P.\ Zendri: Phys.\ Rev.\ Lett.\ {\bf 87},
	031101~(2001)

\bibitem{symon} K.R.\ Symon: \emph{Mechanics}, 2nd edition, chapter 2
	(Addison-Wesley, Reading 1960)

\bibitem{thorne} K.S.\ Thorne: `Gravitational radiation'. In: \emph{300
	Years of Gravitation}, ed.\ by S.W.\ Hawking and W.\ Israel
	(Cambridge University Press, Cambridge 1988), p.\ 330. See
	also more recent reviews by K.S.\ Thorne, e.g.\ 
	{\tt gr-qc/9704042}, or visit the website
	\verb=http://fermi.phys.ualberta.ca/~ccgrra/thorne/index.html=

\bibitem{klm93} A.\ Kr\'olak, J.A.\ Lobo, B.J.\ Meers: Phys.\ Rev.\ D
	{\bf 48}, 3451 (1993)

\bibitem{kay} S.M.\ Kay: \emph{Modern spectral estimation: theory and
	application} (Prentice Hall, New Jersey 1988)

\bibitem{bfs} B.F.\ Schutz: Class.\ Quant.\ Grav.\ {\bf 16}, A131 (1999)

\bibitem{brag} V.B.\ Braginsky, F.Ya.\ Khalili: \emph{Quantum
	measurement} (Cambridge University Press, Cambridge 1995)
\end{thebibliography}
\end{document}